\documentclass[useAmain sequence,usenatbib]{mn2e}
\usepackage{graphicx,subfigure}
\usepackage{bm}
\usepackage{aalongtable}
\usepackage{rotating}

\def\aj{AJ}
\def\an{Astron. Nachr.}
\def\ab{Astrophysical Bulletin}
\def\araa{ARA\&A}
\def\apj{ApJ}
\def\aap{A\&A}
\def\basi{Bull. Astron. Soc. India}
\def\cbet{Cent. Bur. Electron. Telegrams}
\def\jrasc{J. R. Astron. Soc. Canada}
\def\mnras{MNRAS}
\def\pasp{PASP}
\def\rmxaa{Rev. Mexicana Astron. Astrofis.}
\def\sva{Sov.~Astron.}
\def\sval{Sov.~Astron.~Lett.}

\DeclareMathAlphabet{\mathsc}{OT1}{cmr}{m}{sc}
\def\testbx{bx}%
\DeclareRobustCommand{\ion}[2]{%
\relax\ifmmode
\ifx\testbx\f@series
{\mathbf{#1\,\mathsc{#2}}}\else
{\mathrm{#1\,\mathsc{#2}}}\fi
\else\textup{#1\,{\mdseries\textsc{#2}}}%
\fi}
\newcommand{\ha} {\mbox{H$\alpha$}}
\newcommand{\hb} {\mbox{H$\beta$}}
\newcommand{\hg} {\mbox{H$\gamma$}}
\newcommand{\hd} {\mbox{H$\delta$}}
\newcommand{\Feiia} {[\ion{Fe}{ii}]}
\newcommand{\Feii} {\ion{Fe}{ii}}
\newcommand{\Fei} {\ion{Fe}{i}}
\newcommand{\Caiia} {[\ion{Ca}{ii}]}
\newcommand{\Caii} {\ion{Ca}{ii}}

\newcommand{\Baii} {\ion{Ba}{ii}}
\newcommand{\Nai} {\ion{Na}{i}}
\newcommand{\Mgi} {\ion{Mg}{i}}
\newcommand{\Hi} {\ion{H}{i}}
\newcommand{\Hii} {\ion{H}{ii}}
\newcommand{\Scii} {\ion{Sc}{ii}}

\newcommand{\Tiii} {\ion{Ti}{ii}}

\newcommand{\Nii} {\ion{N}{ii}}
\newcommand{\Oia} {[\ion{O}{i}]}
\newcommand{\Oi} {\ion{O}{i}}

\newcommand{\Sii} {\ion{S}{ii}}

\def\sn{SN 2008gz}
\def\gal{NGC 3672}

\newcommand{\eg}{{\textrm e.g.}}
\newcommand{\ie}{{\textrm i.e.}}

\newcommand{\bv}{\mbox{$B\!-\!V$}}
\newcommand{\ebv}{\mbox{$E(B-V)$}}
\newcommand{\bvri}{\mbox{$BVRI$}}

\newcommand{\msun}{\mbox{M$_{\odot}$}}

\newcommand{\kms}{\mbox{$\rm{\,km\,s^{-1}}$}}

\newcommand{\nickel}{\mbox{$^{56}$Ni}}
\newcommand{\cobalt}{\mbox{$^{56}$Co}}
\newcommand{\iron}{\mbox{$^{56}$Fe}}
\newcommand{\mum}{\mbox{$\mu{\rm m}$}}
\newcommand{\el}{\mbox{${e}^{-}$}}

\begin{document}

\title[Supernova 2008gz]
{SN 2008gz $-$ most likely a normal type IIP event}
\author[Roy et al.]
{Rupak Roy$^1$ \thanks{e-mail: roy@aries.res.in, rupakroy1980@gmail.com}, 
 Brijesh Kumar$^{1}$, Alexander S. Moskvitin$^2$, Stefano Benetti$^3$,
\newauthor 
 Timur A. Fatkhullin$^2$, Brajesh Kumar$^{1,4}$, Kuntal Misra$^{5,6}$, 
 Filomena Bufano$^3$,
\newauthor 
 Ralph Martin$^7$, Vladimir V. Sokolov$^2$, S. B. Pandey$^{1,8}$, H. C. Chandola$^9$, 
 Ram Sagar$^1$\\
\\
 $^1$Aryabhatta Research Institute of Observational Sciences (ARIES), Manora 
    Peak, Nainital, 263 129, India\\
 $^2$Special Astrophysical Observatory, Nizhnij Arkhyz, Karachaevo-Cherkesia, 
    369167 Russia\\
 $^3$Istituto Nazionale di Astrofisica, Observatorio Astronomico di Padova, 
    Italy\\
 $^4$Institut d'Astrophysique et de G\'{e}ophysique, 
    Universit\'{e} de Li\`{e}ge, All\'{e}e du 6 Ao\^{u}t 17, B\^{a}t B5c, 
    4000 Li\`{e}ge, Belgium\\
 $^5$Space Telescope Science Institute, 3700 San Martin Drive, Baltimore, MD 
    21218, USA\\
 $^6$Inter University Center for Astronomy and Astrophysics, Post Bag 4, 
    Ganeshkhind, Pune, 411 007, India\\
 $^7$Perth Observatory, 337 Walnut Road, Bickley 6076, Perth, Australia\\
 $^8$Randall Laboratory of Physics, Univ. of Michigan, 450 Church St., Ann 
    Arbor, MI, 48109-1040, USA\\
 $^9$Department of Physics, Kumaun University, Nainital, India\\  
}

\date{Accepted 14-Jan-2011; Received 25-Oct-2010}

\pagerange{\pageref{firstpage}--\pageref{lastpage}} \pubyear{}

\maketitle

\label{firstpage}


\begin{abstract}

 We present $BVRI$ photometric and low-resolution spectroscopic investigation of a 
 type II core-collapse supernova (SN) 2008gz, which occurred in a star forming arm and 
 within a half-light radius (solar metallicity region) of a nearby spiral galaxy NGC 3672. 
 The SN event was detected late, and a detailed investigation of its light curves and spectra 
 spanning 200 days suggest that it is an event of type IIP similar to archetypal 
 SNe 2004et and 1999em. However, in contrast to other events of its class, the SN 2008gz 
 exhibits rarely observed $V$ magnitude drop of 1.5 over the period of a month during 
 plateau to nebular phase. Using 0.21 mag of $A_V$ as a 
 lower limit and a distance of 25.5 Mpc, we estimate synthesized \nickel\, mass 
 of $0.05\pm0.01$\msun\ and a mid-plateau $M_V$ of $-16.6\pm0.2$ mag.
 The photospheric velocity is observed to be higher than
 that was observed for SN 2004et at similar epochs, indicating explosion energy was comparable 
 to or higher than SN 2004et. Similar trend was also seen for the expansion velocity 
 of H-envelopes. By comparing its properties with other well studied events as well as by 
 using a recent simulation of pre-SN models of \cite{dessart10}, we infer an explosion 
 energy range of $2-3\times10^{51}$ erg and this coupled with the observed width of the 
 forbidden \Oia\, 6300-6364\AA\, line at 275 days after the explosion gives an upper 
 limit for the main-sequence (non-rotating, solar metallicity) progenitor mass of 17\,\msun.
 Our narrow-band \ha\, observation, taken nearly 560 days 
 after the explosion and the presence of an emission kink at zero velocity 
 in the Doppler corrected spectra of SN indicate that the event took place in a low 
 luminosity star forming \Hii\, region.

\end{abstract}	



\section{introduction} \label{sec:intro}

 Core-collapse supernovae occur in late type galaxies and their locations are usually 
 associated with the regions of high stellar surface brightness or recent/ongoing star 
 formation, suggesting that they represent the end stages of massive stars 
 ($M >$ 8\,--10 M$_{\sun}$) \citep{anderson09,hakobyan09}. Observationally, these events are 
 classified into H-rich type II SNe which show prominent H-lines in their optical spectra,
 and H-deficient type Ib/c SNe which don't show the trace of H-lines. Ic events lack He-lines as well. 
 Type II SNe constitute about 70\% of all the core-collapse SNe \citep{cappellaro99,smith09} and their
 light curves and spectra differ significantly. In type IIP, the optical light remains constant 
 for about 100 days (called the plateau phase) and then decays exponentially. The spectra are featured 
 due to strong P-cygni profiles, while in type IIL SNe, a linear decline in its optical light and
 strong emission lines are observed. Type IIn events show narrow emission 
 lines \citep{filippenko97,smartt09a}. 

 Theoretically, the explosion mechanism consists of the collapse of the progenitor star's 
 Fe-core, formation of a shock wave, the ejection of stellar envelope and formation of a 
 neutron star or a black hole. 
 The shock wave generated through the reversal of core-collapse, breaks out the stellar 
 surface of progenitor as a hot fireball flashes in X-ray and ultraviolet continuing from a 
 few seconds to a few days. In H-rich events, the shock-heated expanding stellar envelope cools 
 down, by recombination of H and sustains the plateau phase of IIP SNe, while the 
 post-maxima/plateau light curves are powered by the radioactive  decay of \cobalt\, into \iron. 
 Though the explosion mechanism is similar to these events, they differ largely in 
 energetics, e.g. IIP SNe are observed to form a sequence from low-luminosity, low-velocity, 
 Ni-poor events to bright, high-velocity, Ni-rich objects \citep{hamuy03}. Thus, a detailed 
 investigation of individual core-collapse events is important for understanding the nature 
 and environment of progenitors. They generally probe the star formation processes, 
 galactic chemical evolution and constrain the stellar evolutionary 
 models \citep{heger03,smartt09b,habergham10}. Type IIP SNe also turned out to be good 
 standardizable candles \citep{hamuy02,poznanski09,olivares10}. 

 The SN 2008gz event was discovered on November 5.83 UT, 2008 by Koichi 
 Itagaki using a 0.6m telescope in the spiral galaxy NGC 3672 
 at an unfiltered magnitude of about 16.2. On November 7.84 and 8.84 UT,
 an independent discovery of this new transient was reported by R. Martin from 
 Perth Observatory as a part of ``Perth Automated Supernova Search Program'' by using 
 0.6m Lowell Telescope. The red magnitude of this new object 
 was about 15.5 \citep{nakano08}. On November 11.25 UT, \citet{benetti08} 
 took the first spectra of this event with the 3.5m TNG (+ DOLORES; 
 range 340-800 nm, resolution 1.0 nm) and showed that it is a type II supernova
 event and by using GELATO tool \citep{harutyunyan08}, they found that the spectrum of SN 2008gz 
 resembles best with that of a II-peculiar event SN 1998A, taken at 62 days after
 explosion \citep{pastorello05}. 
  Assuming the recession velocity of the host 
 galaxy $\sim$ 1862 \kms, they found the expansion velocity of hydrogen layer was about 
 6600 \kms. An independent regular $BVRI$ CCD photometric monitoring of SN 2008gz was carried out 
 since November 10, 2008 by using 1m Sampurnanand Telescope at Nainital, India.
 We also collected spectra in optical ($0.4-0.9\,\mum$) with 2m IUCAA, India;
 3.6m NTT, Chile; 6m BTA, Russia; 3.6m TNG, Spain. 

 In this work, we present results of optical photometric and low-resolution spectroscopic 
 investigation of SN 2008gz. 
 We adopt time of explosion to be August 20.0, 2009 or JD 2454694.0 having uncertainty of 
 a few days (see \S\ref{sec:phot.curve} for details). Hence the time of post/pre-explosion 
 are rounded off to nearest day and they are referred with $+$ and $-$ signs respectively. 
 Basic properties of SN 2008gz and its host galaxy NGC 3672 are given in Table~\ref{tab:propgal}.
 The paper is organized as follows. \S\ref{sec:phot} and \S\ref{sec:spec} present the 
 photometric and spectroscopic observations and a brief description of light curves
 and spectra. In \S\ref{sec:lines} we study evolution of some important line profiles, where as
 in \S\ref{sec:synow}, the velocity of the photosphere and the H-ejecta is estimated by using 
 SYNOW code \citep{branch01,branch02,baron05} that describes spectroscopic observations.
 Distance, extinction and the evolution of colour
 and bolometric luminosity are studied in \S\ref{sec:dis} and \S\ref{sec:bol} respectively. 
 The amount of synthesized \nickel\, mass, environment and energetics of the progenitor are 
 estimated and discussed in \S\ref{sec:par}. We also made a comparative study of this event with other
 type IIP SNe in \S\ref{sec:comp}. At last a summary is presented in \S\ref{sec:summary}. 



  \begin{table}
  \caption{Properties of the host galaxy NGC 3672 and SN 2008gz.}
  \label{tab:propgal} 
  \centering

  \begin{tabular}{llc} \hline \hline
     \noalign{\smallskip}
      Parameters& Value& Ref.$^{a}$\\ 
     \noalign{\smallskip} \hline
    
     \noalign{\smallskip}
     \multicolumn{3}{l}{\bf NGC 3672:}\\
     Type& SAc& 1\\
     RA (J2000)& $\alpha = 11^{\rm h} 25^{\rm m} 2\fs48$& 1\\
     DEC (J2000)& $\delta = -09\degr 47\arcmin 43\farcs0$& 1\\
     Abs. Magnitude& $M_{B}=-20.59$ mag& 1\\
     \\
     Distance& $D=25.5\pm2.4$ Mpc&  \S\ref{sec:dis}\\
     Scale& $1\arcsec \sim 123$ pc, ~~$1\arcmin \sim 7.4$ kpc&\\
     Distance modulus& $\mu = 32.03\pm0.21$ &\\
     \\
     Apparent radius&${r_{\mathrm{25}}}=1\farcm4\,(\sim 10.4\,\mathrm{kpc})$& 1\\
     Inclination angle&$\Theta_{\rm inc}= 56.2^\circ$&1\\
     Position angle& $\Theta_{\rm maj}=6.5^\circ$&1\\
     Heliocentric Velocity& $cz_{\rm helio}=1864\pm19 \kms$&1\\
     \\
     \multicolumn{3}{l}{\bf SN 2008gz:}\\
     RA (J2000)& $\alpha = 11^{\rm h} 25^{\rm m} 3\fs24$& 2\\
     DEC (J2000)& $\delta = -09\degr 47\arcmin 51\farcs0$& 2\\
     \\
     Location& 13\arcsec\,E, 7\arcsec\,S&  2\\
     Deprojected radius&  $r_{\rm SN} = 23\farcs37$ ($\sim$ 2.81 kpc)& \S\ref{sec:env}\\
     \\
     Explosion epoch (UT)& 20.0 August 2008& \S\ref{sec:phot.curve}\\ 
                   & (JD 2454694.0)& \\ 
     Discovery date (UT)& 5.83 November 2008&  2\\
   
     \noalign{\smallskip}
     \hline
  \end{tabular}
  \newline $^{a}$ (1) HyperLEDA - http://leda.univ-lyon1.fr;
                  (2) \citet{nakano08} 
  
  \end{table}


\begin{figure} 
\centering
\includegraphics[scale = 0.85]{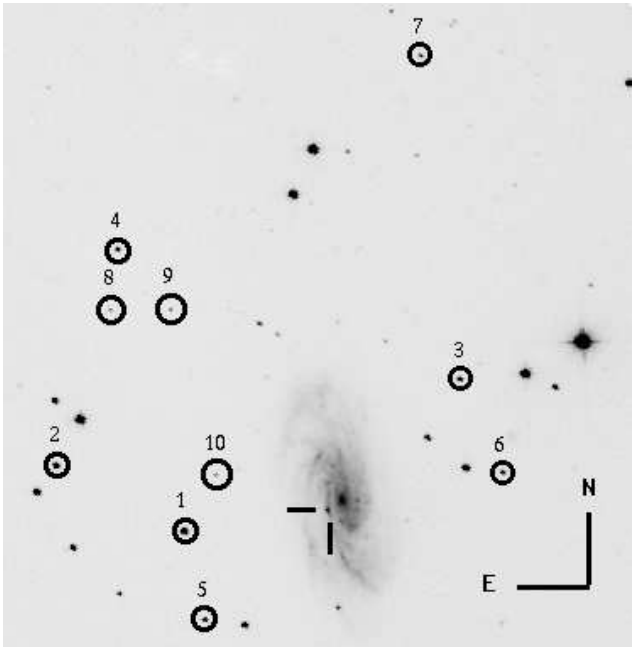}
\caption{SN 2008gz in NGC 3672. $V$ band image from 1 m ST, India.
  Area of about 8 $\times$ 8 arcmin is shown, with location of SN marked with a
  cross and reference standard stars marked with circles.}
\label{fig:snid}
\end{figure}


\section{Broad band photometry} \label{sec:phot}

\subsection{$BVRI$ data} \label{sec:phot.d}

 Initial pre-SN images (--\,403d) of the host galaxy NGC 3672 at $VRI$ bands were 
 obtained from Perth Observatory, as a part of supernova search program for another 
 type Ia SN 2007bm, which occured in the same galaxy. Images were taken with 
 512 $\times$ 512 CCD camera mounted on a 0.6m Lowell Telescope, covering 
 around $5\times5$ square arcmin on the sky. The FWHM seeing was about 2\farcs5.

 SN 2008gz was observed at different epochs from different observatories around
 the world. The major part of monitoring was carried out in Johnson $BV$ and 
 Cousins $RI$ bands from 1m $Sampurnanand$ Telescope (ST)\footnote{ We used a 2048 $\times$ 2048 
 CCD camera having a square pixel of 24\,\mum\,  mounted at the f/13 Cassegrain focus of the 
 telescope. Plate scale of the CCD chip is 0.38 arcsec per pixel, and the entire chip covers 
 a field of 13 $\times$ 13 square arcmin on the sky. The gain and readout noise of the 
 CCD camera are 10 \el\,per analog-to-digital 
 unit and 5.3 \el\,respectively. All the observations were carried out in the binning mode
 of 2$\times$2 pixel.} at the Aryabhatta Research 
 Institute of Observational Sciences (ARIES), Nainital, India.
 SN 2008gz was observed during 10 November 2008 (+87d) to 17 May 2009 (+275d). We could
 not detect SN in the observations of 19 November (+462d) in $B$, of 13 February 2010 (+547d) 
 in $VRI$ and of 14 February 2010 (+548d) in \bvri. 
 In addition to 1m ST, observations of SN 2008gz at \bvri\,bands were also obtained 
 on 21, 22, 24 and 25 March 2009, with IFOSC mounted on 2m IGO, IUCAA, India and on 17 May 2009 
 with EFOSC2 mounted on 3.6m NTT, ESO, Chile. The journal of observations is given in 
 Table~\ref{tab:photlog}\footnote{Table~\ref{tab:photlog} is only available in electronic form.}. 


\begin{table*}
  \caption{Journal of photometric observation of \sn. The full table is available online.
   Please see the supporting information section for detail.} 
  \label{tab:photlog}

  \begin{tabular}{c c c c c c c c cc} \hline
     UT Date&JD&Phase&$B$&$V$&$R$&$I$&Telescope& Seeing& Ellipticity \\
     (yy/mm/dd)&2454000+&(day)&(s)&(s)&(s)&(s)& & (\arcsec)& \\ \hline
     2007/07/09 & 290.94 &$-$403& --           & 240          & 180          & 180          &LT & 2.5& 0.05 \\
     2008/11/10 & 781.48  &$+$87& 2$\times$300 & 2$\times$300 & 2$\times$250 & 2$\times$250 &ST & 2.4& 0.18 \\
     $-$        & $-$     &$-$  & $-$          & $-$          & $-$          & $-$          &$-$& $-$& $-$  \\
     \hline               
  \end{tabular}
  \newline
\end{table*}


 Photometric observations included acquisition of several exposures with 
 exposure time varying from 100 to 300 s in different filters. 
 Several bias and twilight flat frames were obtained for the CCD images. Bias 
 subtraction, flat fielding, cosmic ray removal, alignment and determination of mean FWHM 
 and ellipticity in all the object frames were done using the standard tasks available
 in the data reduction softwares 
 {\it IRAF\,}\footnote{{\it IRAF} stands for Image Reduction and Analysis 
 Facility distributed by the National Optical Astronomy Observatories which 
 is operated by the Association of Universities for research in Astronomy, 
 Inc. under co-operative agreement with the National Science Foundation.}
 and {\it DAOPHOT}\footnote{ {\it DAOPHOT} 
 stands for Dominion Astrophysical Observatory Photometry.} \citep{stetson87,stetson92}.
 The FWHM seeing at $V$ band varied from 2\arcsec\, to 4\arcsec, with a 
 median value of around 2\farcs5. About 10\% of the images taken at large zenith 
 distance had highly elongated PSF (ellipticity $>$ 0.2). For final photometry, we co-added
 the individual frames to increase the signal-to-noise ratio. The pre-processing steps for 
 images taken from other than 1m ST were also performed in similar fashion.

\begin{figure*}
\centering
\includegraphics[scale = 0.65]{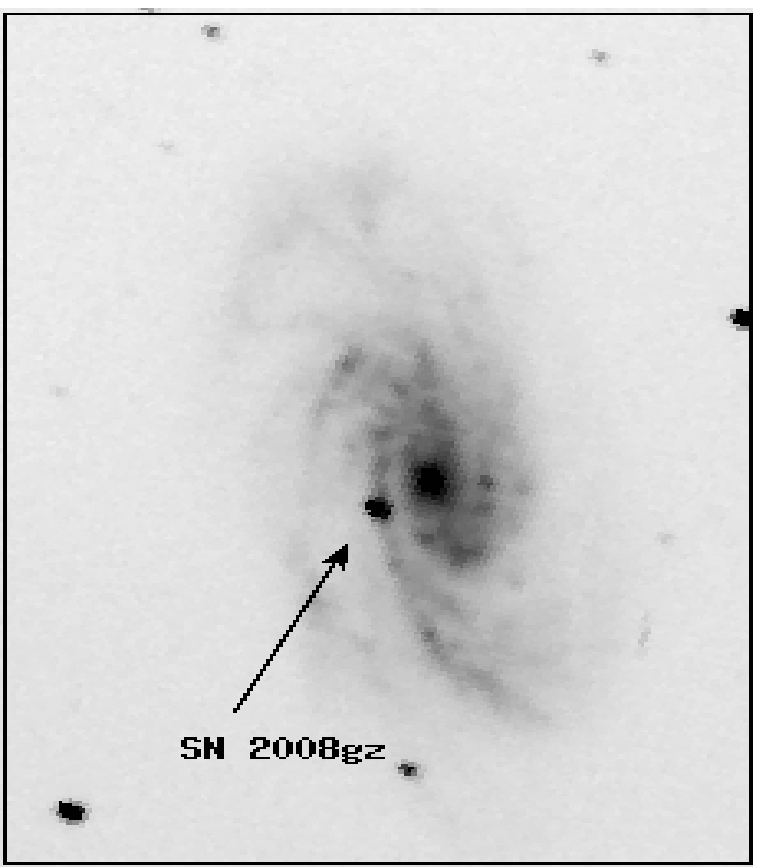}%
\includegraphics[scale = 0.65]{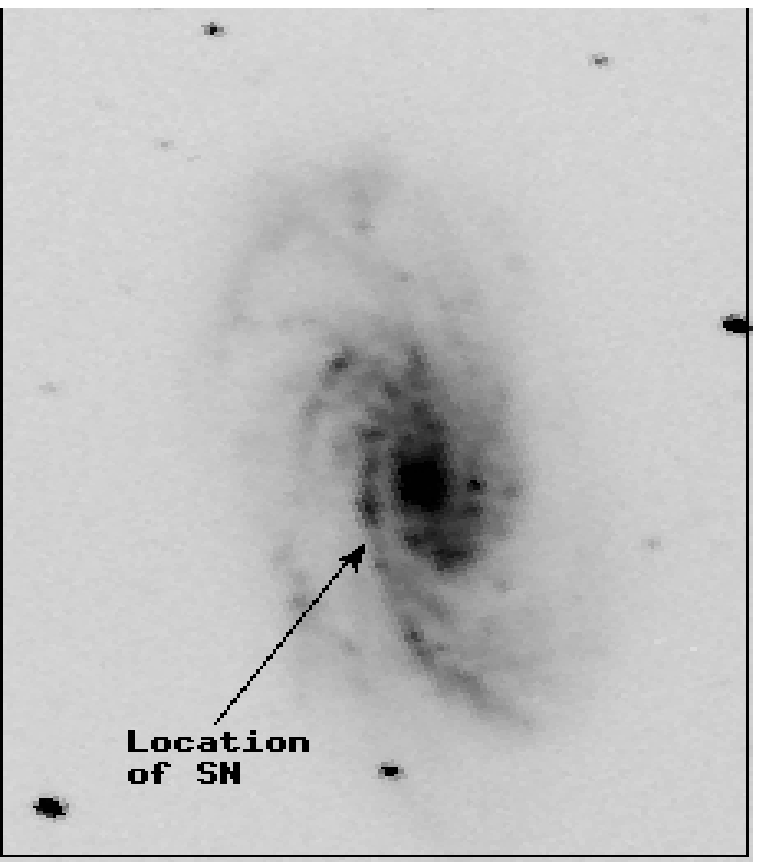}%
\includegraphics[scale = 0.65]{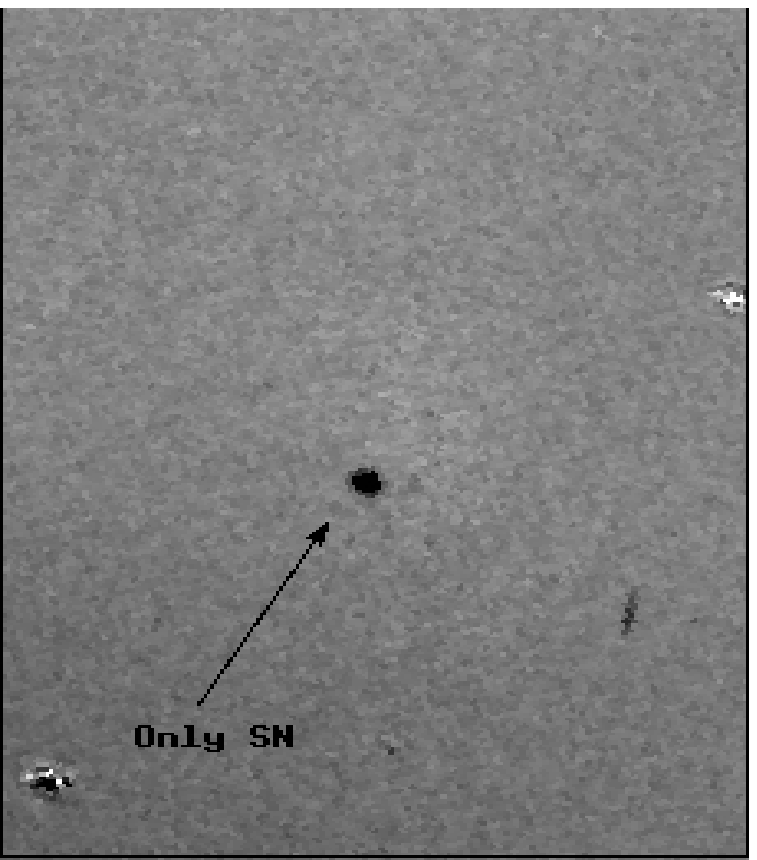}%
\caption{Template subtraction for SN 2008gz. V-band image from 10 December 2008 (+ 117 d) with SN is
 shown in leftmost panel, the template frame from 14 February 2010 (+ 548 d) without SN is shown in middle,
 while the rightmost panel shows the template subtracted image. All the images are around
 $3\times4.5$ arcmin. North is up and East is to left.}
\label{fig:temp}
\end{figure*} 

\begin{table*}
  \caption{Identification number (ID), coordinates ($\alpha, \delta$) and calibrated
           magnitudes of stable secondary standard stars in the field of SN 2008gz. Errors in
           magnitude represent RMS scatter in the night-to-night repeatability  
           over entire period of SN monitoring.}
  \label{tab:photstar}

  \begin{tabular}{ccccccc}
     \hline
     Star& $\alpha_{\rm J2000}$&       $\delta_{\rm J2000}$&  $B$&  $V$&  $R$&  $I$\\
       ID&          (h m s)&(\degr\,\arcmin\,\arcsec)&(mag)&(mag)&(mag)&(mag)\\
     \hline
      1&11 25 11.27&-09 48 04.4&15.36$\pm$0.01&14.36$\pm$0.03&13.99$\pm$0.01&13.49$\pm$0.01\\
      2&11 25 18.34&-09 47 03.0&15.70$\pm$0.01&14.92$\pm$0.02&14.68$\pm$0.01&14.25$\pm$0.01\\
      3&11 24 55.62&-09 46 07.9&16.50$\pm$0.01&15.82$\pm$0.03&15.64$\pm$0.01&15.26$\pm$0.01\\
      4&11 25 14.22&-09 44 04.7&16.91$\pm$0.01&16.15$\pm$0.01&15.93$\pm$0.01&15.50$\pm$0.01\\
      5&11 25 10.47&-09 49 17.0&17.00$\pm$0.02&16.23$\pm$0.03&15.98$\pm$0.01&15.52$\pm$0.04\\
      6&11 24 53.17&-09 47 25.1&17.09$\pm$0.02&16.42$\pm$0.03&16.22$\pm$0.01&15.83$\pm$0.03\\
      7&11 24 56.67&-09 41 35.1&17.80$\pm$0.02&16.99$\pm$0.02&16.66$\pm$0.01&16.19$\pm$0.04\\
      8&11 25 14.78&-09 44 54.3&18.50$\pm$0.07&17.70$\pm$0.07&17.49$\pm$0.06&17.12$\pm$0.03\\
      9&11 25 11.40&-09 44 54.5&18.28$\pm$0.04&17.75$\pm$0.03&17.62$\pm$0.02&17.22$\pm$0.04\\
     10&11 25 09.31&-09 47 16.4&18.83$\pm$0.09&17.92$\pm$0.07&17.63$\pm$0.03&17.08$\pm$0.06\\
     \hline
  \end{tabular}

\end{table*}


 Fig.~\ref{fig:snid} shows the location of \sn\, in the galaxy \gal. The SN flux is expected 
 to have substantial contribution from the host galaxy background, due to its proximity 
 to the galaxy centre, its location in a spiral arm and a high inclination 
 angle (56.2\degr; Table~\ref{tab:propgal}) of the galaxy. 
 At early phases, SN flux dominates the total flux, thus with a PSF-fitting method
 we were able to remove the galaxy contribution. At later epochs (\eg\, end of
 plateau in case of type IIP events), galaxy flux may brighten the
 SN light curves by 0.5 to 1 mag depending on its location in the galaxy \citep{pastorello05}. 
 We used {\it ISIS}\footnote{http://www2.iap.fr/users/alard/package.html} \citep{alard98} 
 to get galaxy template subtracted flux of supernova. 
 As a template we used the \bvri\, images taken on 14 February 
 2010 (+548d) from 1m ST, India in good seeing conditions. We note that the galaxy 
 subtraction using pre-SN (--403d) $VRI$ images taken 
 from Perth Observatory, gave no SN contribution above noise level in our images
 recorded on 19 November 2009, 13 and 14 February 2010. In order to verify
 ISIS results, we also performed the galaxy template subtraction scheme independently
 using self-written scripts employing IRAF tasks which included alignment,
 PSF and intensity matching of the galaxy template and SN images, and subtraction
 of template from SN images. Fig.~\ref{fig:temp} shows images with and without 
 template subtraction. PSF-fitting method was applied on the subtracted images. 
 Our magnitude was found to be consistent with the ISIS ones having a typical scatter 
 of $\sim$0.1 mag in $BVRI$; this scatter is of the order of 0.05 mag in plateau phase
 (see Fig.~\ref{fig:diffplot}\footnote{Figure~\ref{fig:diffplot} 
 is only available in electronic form.}).

 In order to calibrate instrumental magnitudes of SN 2008gz, we observed
 \citet{landolt09} standard fields SA92 and PG0231 in $BVRI$ with 1m ST on 
 15 November 2008 under good night conditions 
 (transparent sky, FWHM seeing in $V \sim 2\arcsec$). The data reduction
 of SN and Landolt fields were done using profile fitting technique and
 the instrumental magnitudes were converted into standard system following least-square
 linear regression procedures outlined in \citet{stetson92}. We used mean
 values of atmospheric extinction coefficients of the site viz. 0.28,
 0.17, 0.11 and 0.07 mag per unit airmass for the $B$, $V$, $R$ and $I$ 
 bands respectively \citep{kumar00}. A set of 13 stars having a colour range of 
 $-0.33 \leq (\bv) \leq 1.45$ and brightness range of $12.77 \leq V \leq 16.11$ 
 were used to derive the following zero points and colour coefficients:

 \[ b = B + (5.37\pm0.02) + (0.01\pm0.02)(B-V) \]
 \[ v = V + (4.89\pm0.01) +(-0.02\pm0.02)(B-V) \]
 \[ r = R + (4.72\pm0.01) + (0.02\pm0.02)(V-R) \]
 \[ i = I + (5.17\pm0.02) + (0.02\pm0.02)(V-I) \]

 \noindent
 Here $B$, $V$, $R$, $I$ are the standard magnitudes and  $b$, $v$, $r$, $i$ are 
 corresponding instrumental magnitudes corrected for time and aperture. A typical scatter
 in the photometric solutions to the Landolt standard stars is found to be $\sim$ 0.03 mag 
 for $BVRI$. Table~\ref{tab:photstar}
 lists the calibrated magnitudes for a set of ten stable secondary standards in 
 the SN field, while calibrated \bvri magnitudes of \sn\, are presented in 
 Table~\ref{tab:photsn}. For SN, we quote ISIS derived errors ($1\sigma$ uncertainty), 
 which is consistent with the RMS scatter in the magnitude of standard stars determined 
 from night-to-night repeatability over entire period ($\sim$ 215d) of SN monitoring. 
 Large errors in 2m IGO and 3.6m NTT data arises due to mismatch in the PSF and pixel scale.

\setcounter{figure}{3}
\begin{figure}
\centering
\includegraphics[scale = 0.415]{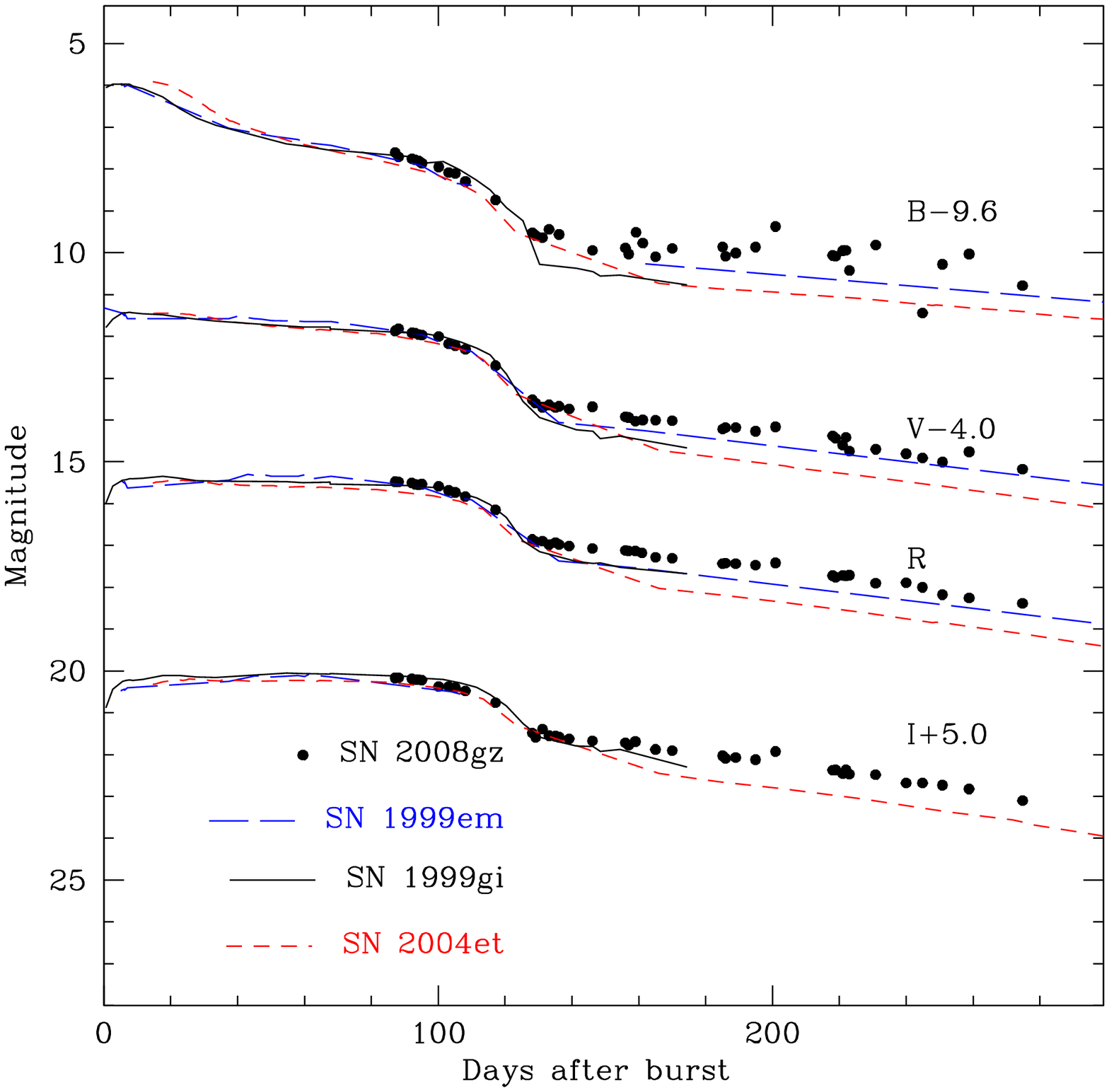}
\caption{Light curve in $BVRI$ magnitudes of SN 2008gz. The light curves are shifted
  for clarity, while for other SNe, they are scaled in magnitude and time to match with SN 2008gz.}
\label{fig:lightcur}
\end{figure}

\subsection{Optical light curve} \label{sec:phot.curve}

 Fig.~\ref{fig:lightcur} shows $BVRI$ light curves of SN 2008gz ranging 
 from +87d to +275d since the time of explosion. We also present
 the light curves of other well studied nearby ($D < 12\,{\rm Mpc}$) 
 type IIP SNe viz. 2004et \citep{sahu06,misra07}, 1999em \citep{elmhamdi03a} 
 and 1999gi \citep{leonard02}, scaled in time and magnitude to match the transition 
 between plateau to nebular phases. It is seen that SN 2008gz was detected close to 
 the end of its plateau phase and its light curve resembles well with the above 
 three template IIP events and hence we could determine (and adopt) the time of inflection
 (plateau to nebular), $t_{i}$ of $115\pm5$ d, by adjusting the template light curves to get 
 the best match to \sn\, data points. This derived plateau duration is typical for type II P events 
 \citep{elmhamdi03b} and it is also consistent with the fact that SN 2008gz was not 
 visible around three months before the discovery date (November 5.83, 2008) at the 
 level of unfiltered magnitude of 19.0 \citep{nakano08}. Further, similarities between 
 bolometric tail luminosity of SN 2008gz (see \S\ref{sec:bol}) with that of 
 SN 2004et and SN 1999em indicates that probably explosion happened about 
 82 days before the discovery date. Analysis of first spectrum of SN 2008gz 
 (see \S\ref{sec:spec.spectra}) also revealed few similarities between the kinematical 
 properties of its ejecta to that of SN 2004et observed nearly 80 days after the 
 burst. We therefore adopt time of SN explosion to be $82\pm5$ days before
 the discovery date and this corresponds to burst time, $t_{0}$ of JD 2454694.0,
 however, we note that based on the first spectrum (November 11.25 UT, 2008) and its 
 similarity with +62d spectrum of SN 1998A \citep{benetti08} suggesting a time of explosion 
 of nearly 56 days before the discovery date (corresponding 
 to plateau phase of $\sim$ 90d) cannot be ruled out.

 In the late plateau phase ($\sim$ +90d) flatness behaviour in $RI$ and decline trend in $BV$ 
 are clearly seen, which are similar to other IIP events. The $V$ magnitude drop of 1.5 mag from
 plateau phase ($V \sim$ 16 mag at +100d) to nebular phase (17.5 mag at +130d),  
 is slightly lower than 2-3 mag drop for a typical IIP 
 event \citep{olivares10}. This shallow decline which is seen in $BRI$ also, 
 indicating production of large \nickel\, mass (see \S\ref{sec:nick}). In contrary
 to this, very steep brightness decline at $V$ has also been observed, e.g. 4.5 mag 
 for SN 2007od \citep{andrews10}. 
 The nebular phase starts at $\sim$ +140d, and it roughly 
 follows the decay slope of \cobalt\, to \iron : 0.98 mag $(100 {\rm d})^{-1}$. 
 A linear fit to the tail from +150d to +275d gives the following decline rates
 [in mag $(100{\rm d})^{-1}$]: $\gamma_{B} \sim 0.51$, $\gamma_{V} \sim 0.98$, 
 $\gamma_{R} \sim 1.12$, $\gamma_{I} \sim 1.13$  at $B$, $V$, $R$, $I$ which is typical to the 
 values found for IIP SNe. The flattening seen in $B$ band light curve, though 
 non-conclusive due to large scatter of the measurements, has also been observed in other
 events, e.g. in 1999em \citep{elmhamdi03a} and 1987A \citep{suntzeff88} until +400d.


\section {Low resolution spectroscopy} \label{sec:spec}

\subsection{Data} \label{sec:spec.d}

 Long-slit low resolution spectra ($\sim$ 6 to 14 \AA) in 
 the optical range (0.33 - 1.0 \mum) were collected at eight epochs during +87d to +275d; 
 five epochs from 2m IGO, and one epoch each from 3.5m TNG, 6m BTA and 3.6m NTT. 
 Journal of spectroscopic observations are given in Table~\ref{tab:speclog}. 
 
 At 2m IGO, observations
 were carried out using IFOSC (IUCAA Faint Object Spectrograph and Camera) mounted
 at the cassegrain end of f/10 reflector \citep{gupta02,chakraborty05}. Slit spectra
 were recorded using $2048\times2048$ EEV CCD camera with 13.5 \mum\, pixel, having a gain 
 of 1.8 \el per analog-to-digital unit, and readout noise of 6.3 \el. 
 Grism 7 with peak sensitivity at 500 nm and
 a slit width of 1\farcs5 were used. Calibration frames (bias, flats, arcs) and 
 spectrophotometric flux standards were observed on each night. 
 For SN, usually slits were placed across the spiral arm so as to make proper sky 
 background and in one case at +170d, the galaxy centre was also observed.    
 Spectroscopic data reduction was done under IRAF environment. Bias and flatfielding were
 performed on each frames. Cosmic ray rejection on each frame was done by using 
 Laplacial kernel detection \citep{dokkum01}. Images were coadded to improve the
 signal-to-noise ratio and one-dimensional spectra were extracted using {\it apall} 
 task in IRAF which is based on optimal extraction algorithm by \citet{horne86}. 
 Wavelength calibration were performed by using {\it identify} task and about 15-18 emission 
 lines of He and Ar which were used to find a dispersion solution. Fifth order fits were used
 to achieve a typical RMS uncertainty of 0.1\AA. The position of OI emission skyline 
 at 5577\AA\, was used to check the wavelength calibration and deviations were found
 between 0.5 to 1\AA\, and this was corrected by linear shift in dispersion. 
 The instrumental FWHM resolution of 2m IGO spectra as measured from \Oi\,5577\AA\, 
 emission skyline was found to lie between 6\AA\, to 10\AA\, ($\sim$ 322 - 510 \kms). 
 Flux calibration was done using spectrophotometric fluxes from \citet{hamuy94} 
 and assuming a mean extinction for the site. Synthetic magnitudes were estimated using
 spectra to verify the accuracy of flux calibration and it was found to be accurate
 within 0.05 mag. 

 Spectroscopic data reduction for DOLORES on 3.6m TNG, EFOSC2 on 3.6m NTT, and SCORPIO 
 on 6m BTA were done in similar fashion and at around 6000\AA\, it had a resolution of
 10\AA, 14\AA\, and 12\AA\, respectively. 

\subsection{Optical spectra} \label{sec:spec.spectra}

\begin{figure}
\centering
\includegraphics[scale = 0.45]{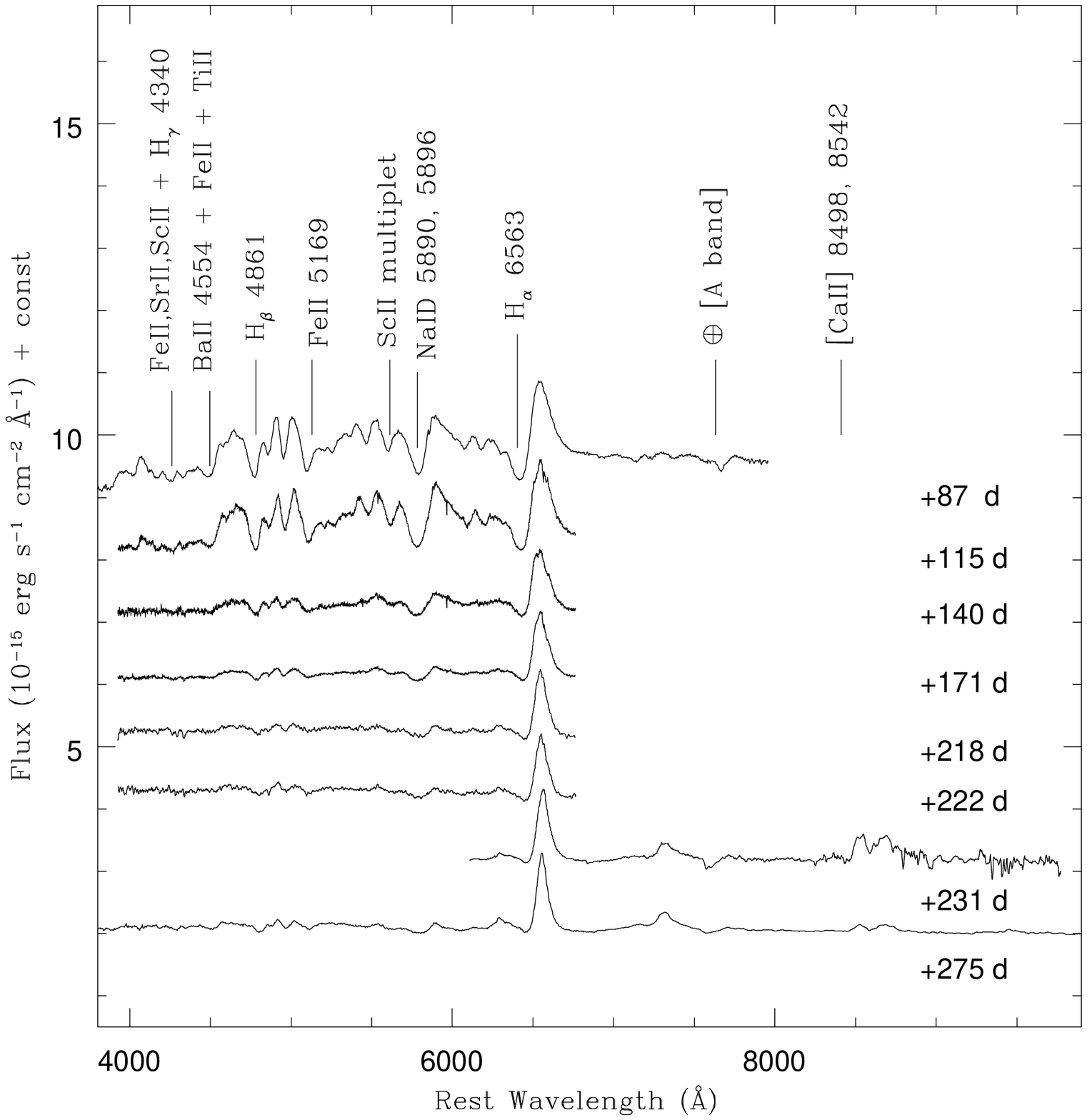}
\caption{Doppler corrected flux spectra of SN 2008gz from late plateau (+87 d) to
         nebular phase (+275 d). Prominent Hydrogen and metal lines are marked.}
\label{fig:specall}
\end{figure}

\begin{figure*}
\centering
\includegraphics[scale = 0.73]{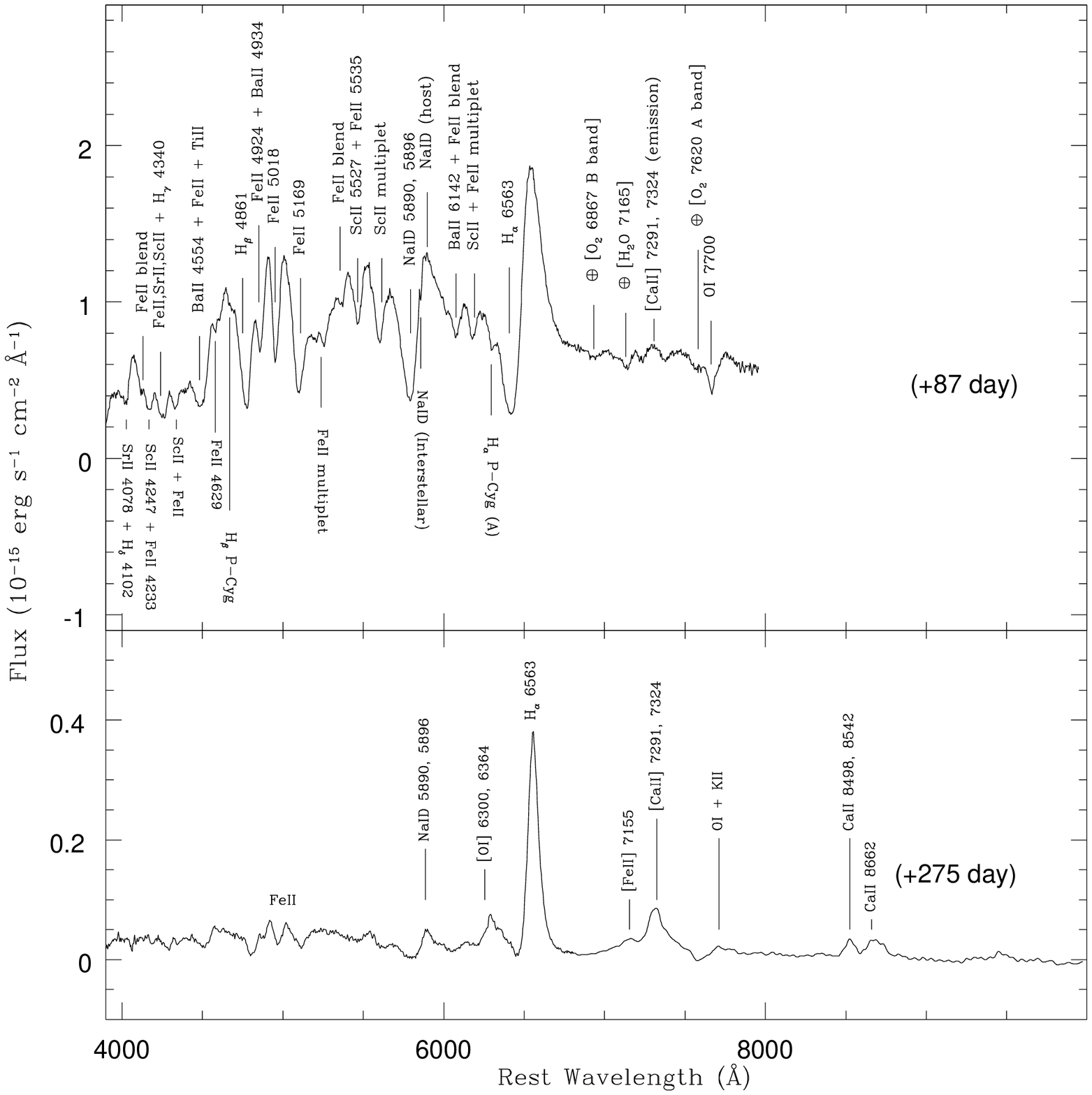}%
\caption{Line identification for late plateau and deep nebular phases.}
\label{fig:specid}
\end{figure*}

 Fig.~\ref{fig:specall} and Fig.~\ref{fig:specid} show the rest frame spectra of \sn\,, corrected 
 for recession velocity (1864 \kms) of the host galaxy NGC 3672. We identify all the 
 spectral features as per previously published line identifications for IIP events 
 \citep{leonard02, sahu06}. In Fig.~\ref{fig:specall}, end of the plateau 
 phase (+115d) and beginning of nebular phase (+140d) is clearly evident in the 
 spectral evolution. 

\begin{table*}
  \caption{Photometric evolution of SN 2008gz. Errors in magnitude are derived from ISIS and it 
    denote $1\sigma$ uncertainty.}
  \label{tab:photsn}
  \begin{tabular}
  {c c c c c c c c c}
  \hline
  UT Date&JD&Phase$^{a}$&$B$&$V$&$R$&$I$&Telescope$^{b}$& Seeing$^{c}$ \\
  (yy/mm/dd)&2454000+&(day)&(mag)&(mag)&(mag)&(mag)&& (\arcsec)\\
  \hline
  2008/11/10 & 781.47 &  $+$87& 17.205$\pm$0.016 & 15.871$\pm$0.012 & 15.475$\pm$0.008 & 15.168$\pm$0.015 &ST  & 2.4 \\
  2008/11/11 & 782.49 &  $+$88& 17.312$\pm$0.030 & 15.819$\pm$0.014 & 15.477$\pm$0.010 & 15.164$\pm$0.013 &ST  & 2.6 \\
  2008/11/15 & 786.47 &  $+$92& 17.355$\pm$0.040 & 15.918$\pm$0.019 & 15.501$\pm$0.013 & 15.189$\pm$0.017 &ST  & 2.7 \\
  2008/11/16 & 787.47 &  $+$93& 17.376$\pm$0.033 & 15.920$\pm$0.017 & 15.537$\pm$0.011 & 15.210$\pm$0.014 &ST  & 2.4 \\
  2008/11/17 & 788.48 &  $+$94& 17.402$\pm$0.030 & 15.957$\pm$0.017 & 15.552$\pm$0.011 & 15.207$\pm$0.016 &ST  & 2.4 \\
  2008/11/18 & 789.46 &  $+$95& 17.457$\pm$0.037 & 15.966$\pm$0.021 & 15.533$\pm$0.012 & 15.219$\pm$0.015 &ST  & 3.3 \\
  2008/11/23 & 794.47 & $+$100& 17.550$\pm$0.033 & 16.002$\pm$0.020 & 15.587$\pm$0.015 & 15.373$\pm$0.019 &ST  & 3.5 \\
  2008/11/26 & 797.47 & $+$103& 17.684$\pm$0.014 & 16.182$\pm$0.008 & 15.685$\pm$0.003 & 15.373$\pm$0.005 &ST  & 2.7 \\
  2008/11/28 & 799.45 & $+$105& 17.703$\pm$0.029 & 16.228$\pm$0.018 & 15.732$\pm$0.014 & 15.397$\pm$0.021 &ST  & 3.0 \\
  2008/12/01 & 802.49 & $+$108& 17.900$\pm$0.028 & 16.311$\pm$0.017 & 15.830$\pm$0.011 & 15.475$\pm$0.017 &ST  & 2.5 \\
  2008/12/10 & 811.49 & $+$117& 18.342$\pm$0.037 & 16.702$\pm$0.027 & 16.151$\pm$0.017 & 15.761$\pm$0.026 &ST  & 2.9 \\
  2008/12/21 & 822.49 & $+$128& 19.125$\pm$0.094 & 17.519$\pm$0.047 & 16.857$\pm$0.026 & 16.480$\pm$0.040 &ST  & 2.6 \\
  2008/12/22 & 823.46 & $+$129& 19.190$\pm$0.073 & 17.597$\pm$0.043 & 16.909$\pm$0.023 & 16.583$\pm$0.029 &ST  & 2.3 \\
  2008/12/24 & 825.53 & $+$131& 19.242$\pm$0.174 & 17.695$\pm$0.058 & 16.893$\pm$0.022 & 16.393$\pm$0.026 &ST  & 3.6 \\
  2008/12/26 & 827.52 & $+$133& 19.044$\pm$0.176 & 17.638$\pm$0.036 & 16.982$\pm$0.017 & 16.546$\pm$0.024 &ST  & 3.1 \\
  2008/12/28 & 829.44 & $+$135&     --           & 17.704$\pm$0.037 & 16.928$\pm$0.020 & 16.559$\pm$0.026 &ST  & 2.7 \\
  2008/12/29 & 830.49 & $+$136& 19.165$\pm$0.085 & 17.674$\pm$0.039 & 16.976$\pm$0.021 & 16.582$\pm$0.041 &ST  & 3.2 \\
  2009/01/01 & 833.52 & $+$139&     --           & 17.738$\pm$0.047 & 17.016$\pm$0.021 & 16.623$\pm$0.030 &ST  & 2.7 \\
  2009/01/08 & 840.48 & $+$146& 19.547$\pm$0.121 & 17.684$\pm$0.043 & 17.069$\pm$0.024 & 16.671$\pm$0.049 &ST  & 3.1 \\
  2009/01/18 & 850.36 & $+$156& 19.489$\pm$0.168 & 17.920$\pm$0.082 & 17.118$\pm$0.040 & 16.717$\pm$0.055 &ST  & 3.1 \\
  2009/01/19 & 851.32 & $+$157& 19.631$\pm$0.156 & 17.941$\pm$0.068 & 17.128$\pm$0.033 & 16.779$\pm$0.061 &ST  & 3.1 \\
  2009/01/21 & 853.48 & $+$159& 19.115$\pm$0.069 & 18.027$\pm$0.078 & 17.128$\pm$0.038 & 16.688$\pm$0.049 &ST  & 3.2 \\
  2009/01/23 & 855.51 & $+$161& 19.373$\pm$0.077 & 18.000$\pm$0.102 & 17.178$\pm$0.036 &   --             &ST  & 3.1 \\
  2009/01/27 & 859.36 & $+$165& 19.699$\pm$0.096 & 18.002$\pm$0.054 & 17.279$\pm$0.029 & 16.877$\pm$0.045 &ST  & 2.2 \\
  2009/02/01 & 864.35 & $+$170& 19.498$\pm$0.093 & 18.019$\pm$0.062 & 17.305$\pm$0.032 & 16.903$\pm$0.047 &ST  & 2.5 \\
  2009/02/16 & 879.42 & $+$185& 19.461$\pm$0.096 & 18.217$\pm$0.065 & 17.434$\pm$0.032 & 17.029$\pm$0.046 &ST  & 2.4 \\
  2009/02/17 & 880.27 & $+$186& 19.684$\pm$0.095 & 18.181$\pm$0.064 & 17.420$\pm$0.030 & 17.087$\pm$0.046 &ST  & 2.9 \\
  2009/02/20 & 883.34 & $+$189& 19.609$\pm$0.124 & 18.179$\pm$0.075 & 17.434$\pm$0.039 & 17.071$\pm$0.058 &ST  & 3.2 \\
  2009/02/26 & 889.27 & $+$195& 19.467$\pm$0.113 & 18.269$\pm$0.093 & 17.468$\pm$0.045 & 17.121$\pm$0.068 &ST  & 3.6 \\
  2009/03/04 & 895.23 & $+$201& 18.976$\pm$0.118 & 18.163$\pm$0.111 & 17.417$\pm$0.052 & 16.924$\pm$0.070 &ST  & 3.0$^{d}$ \\
  2009/03/21 & 912.33 & $+$218& 19.665$\pm$0.877 & 18.380$\pm$0.341 & 17.720$\pm$0.125 & 17.373$\pm$0.181 &IGO & 1.6 \\
  2009/03/22 & 913.24 & $+$219& 19.681$\pm$0.775 & 18.441$\pm$0.335 & 17.762$\pm$0.128 & 17.366$\pm$0.196 &IGO & 1.6 \\
  2009/03/24 & 915.32 & $+$221& 19.552$\pm$0.712 & 18.601$\pm$0.292 & 17.714$\pm$0.104 & 17.449$\pm$0.165 &IGO & 1.5 \\
  2009/03/25 & 916.30 & $+$222& 19.545$\pm$0.727 & 18.416$\pm$0.320 & 17.723$\pm$0.117 & 17.362$\pm$0.171 &IGO & 1.4 \\
  2009/03/26 & 917.32 & $+$223& 20.029$\pm$0.188 & 18.743$\pm$0.143 & 17.710$\pm$0.057 & 17.463$\pm$0.107 &ST  & 3.5 \\
  2009/04/03 & 925.20 & $+$231& 19.414$\pm$0.372 & 18.701$\pm$0.242 & 17.899$\pm$0.083 & 17.481$\pm$0.130 &ST  & 2.8 \\
  2009/04/12 & 934.29 & $+$240&    --            & 18.807$\pm$0.257 & 17.888$\pm$0.089 & 17.680$\pm$0.124 &ST  & 2.7 \\
  2009/04/17 & 939.23 & $+$245& 21.040$\pm$0.270 & 18.909$\pm$0.132 & 18.000$\pm$0.052 & 17.680$\pm$0.080 &ST  & 2.8 \\
  2009/04/23 & 945.16 & $+$251& 19.878$\pm$0.187 & 19.005$\pm$0.165 & 18.176$\pm$0.093 & 17.730$\pm$0.145 &ST  & 3.1 \\
  2009/05/01 & 953.15 & $+$259& 19.633$\pm$0.182 & 18.760$\pm$0.199 & 18.254$\pm$0.080 & 17.822$\pm$0.109 &ST  & 2.5 \\
  2009/05/17 & 969.11 & $+$275& 20.388$\pm$0.984 & 19.173$\pm$0.556 & 18.383$\pm$0.269 & 18.099$\pm$0.281 &NTT & 0.9 \\
  \hline                                                                                                          
  \end{tabular}                                                                                                   
  \newline\newline                                                                                                
  $^{a}$ with reference to the explosion epoch JD 2454694.0\\
  $^{b}$ ST : 1 m Sampurnanand Telescope, ARIES, India; IGO : 2 m IUCAA 
  Girawali Observatory, IUCAA, India; NTT : 3.6 m New Technology Telescope, ESO, Chile\\
  $^{c}$ FWHM of the stellar PSF at $V$ band\\
  $^{d}$ Flat field problem
\end{table*}


 The late plateau phase (+87d and +115d) spectra are marked 
 by strong P-Cygni features of \ha, \Oi\,7700\AA, 
 \Nai\,D 5890, 5896\AA, and singly ionised Sc, Ba, Ti, Fe atoms, 
 while the +140d and later spectra show significant drop in the absorption 
 strength of P-Cygni features. The spectra during +115d to +222d show the spectral 
 evolution of the event from early to mid-nebular stage, while the last two 
 spectra (+231d and +275d) are those typical shown during the late stages of a typical
 SNIIP. In Fig.~\ref{fig:specid}, the +87d spectrum shows various atomic 
 absorption lines over the weak continuum. These lines are mainly
 due to elements present in the SN ejecta along with some earth atmospheric 
 molecular lines (marked with $\oplus$) and absorption due \Nai\,D of Milky Way 
 and host galaxy. On the other hand, the +275d  spectrum shows a typical
 nebular phase spectrum dominated by emission lines.


\begin{table*}
  \caption{Journal of spectroscopic observations of SN 2008gz.}
  \label{tab:speclog}
  \begin{tabular}{lc cc cc cc cc}
    \hline
    UT Date &JD&Phase$^{a}$&    Range&Telescope$^{b}$& Grating& Slit width& Dispersion& Exposure&S/N$^{c}$ \\
    (yy/mm/dd/hh.hh) &2454000+&(days)& \mum&      & (gr mm$^{-1}$)&  (\arcsec)& (\AA\,pix$^{-1}$)& (s)&(pix$^{-1}$)\\ \hline

        2008/11/11/06.049& 781.88&  +87& 0.34$-$0.80& TNG& 500& 1.5 & 2.5&       1200& 80\\
    
        2008/12/08/23.346& 809.46& +115& 0.38$-$0.68& IGO& 600& 1.5 & 1.4&     2x1800&50\\
        2009/01/03/00.333& 834.52& +140& 0.38$-$0.68& IGO& 600& 1.5 & 1.4&     2x1800&26\\
  2009/02/01/22.208$^{d}$& 864.40& +170& 0.38$-$0.68& IGO& 600& 1.5 & 1.4&      2x1800&20\\
        2009/02/02/21.400& 865.40& +171& 0.38$-$0.68& IGO& 600& 1.5 & 1.4&     3x1800&24\\
        2009/03/22/18.303& 913.27& +218& 0.38$-$0.68& IGO& 600& 1.5 & 1.4&       1800&8\\
        2009/03/25/19.673& 916.33& +222& 0.38$-$0.68& IGO& 600& 1.5 & 1.4&       1800&7\\
    
        2009/04/03/22.643& 924.50& +231& 0.61$-$1.00& BTA& 550& 2.1 & 3.5&      3x900  &70\\
    
        2009/05/17/00.543& 969.04& +275& 0.33$-$0.80& NTT& 300& 1.0 & 4.0&       2700&16\\
        2009/05/17/01.308& 969.07& +275& 0.55$-$1.05& NTT& 300& 1.0 & 4.2&       2700&24\\
    \hline
  \end{tabular}
  \newline
  $^{a}$ With reference to the burst time JD 2454694.0\\
  $^{b}$ TNG : DOLORES on 3.5m Telescopio Nazionale Galileo, Italy; IGO : IFOSC on
  2m IUCAA Girawali Observatory, India; BTA : SCORPIO on 6m Big Telescope Alt-azimuthal,
  Special Astrophysical Observatory, Russia; 
  NTT : EFOSC2 on 3.6m New Technology Telescope, ESO, Chile. \\
  $^{c}$ At 0.6 \mum\\
  $^{d}$ Only center of galaxy observed 
\end{table*}


 Temporal evolution of P-Cygni nature of \ha\, 
 is clearly seen viz. the emission component becomes narrower with a 
 decrease in depth of associated absorption component during the transition 
 of SN from plateau to nebular phase. The FWHM of emission component of \ha\ 
 decreases from $\sim5477\kms$ at +87d to $\sim3526\kms$ at $+275$d, indicating
 decrease in opacity and temperature of \Hi\, line emitting regions. For \hb, \hg\, 
 and \hd, the emission components are crowded with numerous metal lines.
 In Fig.~\ref{fig:specid}, we also see impression of an 
 additional P-Cygni component in the absorption profile of \ha\, and \hb. 
 This is speculated as a footprint of high velocity emitting shells in SN ejecta.
 Similar signatures were also noticed in type IIP SNe 1999em and 
 2004et \citep{leonard02,li05,sahu06}. 

 The spectrum, labeled with $+275$d shows a typical late nebular phase spectrum marked by
 emission dominated permitted lines of 
 \Caii\, 8498, 8542, 8662\AA\, and \Nai\,D as well as with 
 the appearance of forbidden emission lines, \ie, \Oia\,6300, 6364\AA;
 \Feiia\,7155\AA\, and \Caiia\,7291, 7324\AA\,. These forbidden lines are not observed at Earth due 
 to high density of gas. \Caiia\, is already seen in +87d spectrum, \Oia\, appears at
 +171d, while the \Feiia\, appears in +275d spectrum. Increasing strength of these 
 forbidden lines indicate  expansion and rarefaction of SN ejecta with time. 
 The \Feiia\, line is also noticed 
 by \citet{pastorello05} in +344d spectrum of SN 1998A and +346d spectrum of SN 1987A, 
 whereas for low luminosity SN 1997D it was visible at +417d \citep{benetti01}. We 
 determine relative strength $I(\Caiia)/I(\Feiia)$ of 2.06 for \sn, whereas for 
 SNe 1999em and 1987A, values of this ratio are respectively 6.98 and 24.68 at around 
 +400d \citep{elmhamdi03a}. This indicates that the physical conditions of \Feiia\
 formation in \sn\, may be similar to that of SN 1999em rather than SN 1987A.

\section{Temporal evolution of spectral lines} \label{sec:lines}

\begin{figure*}
\centering
\includegraphics[scale = 0.77]{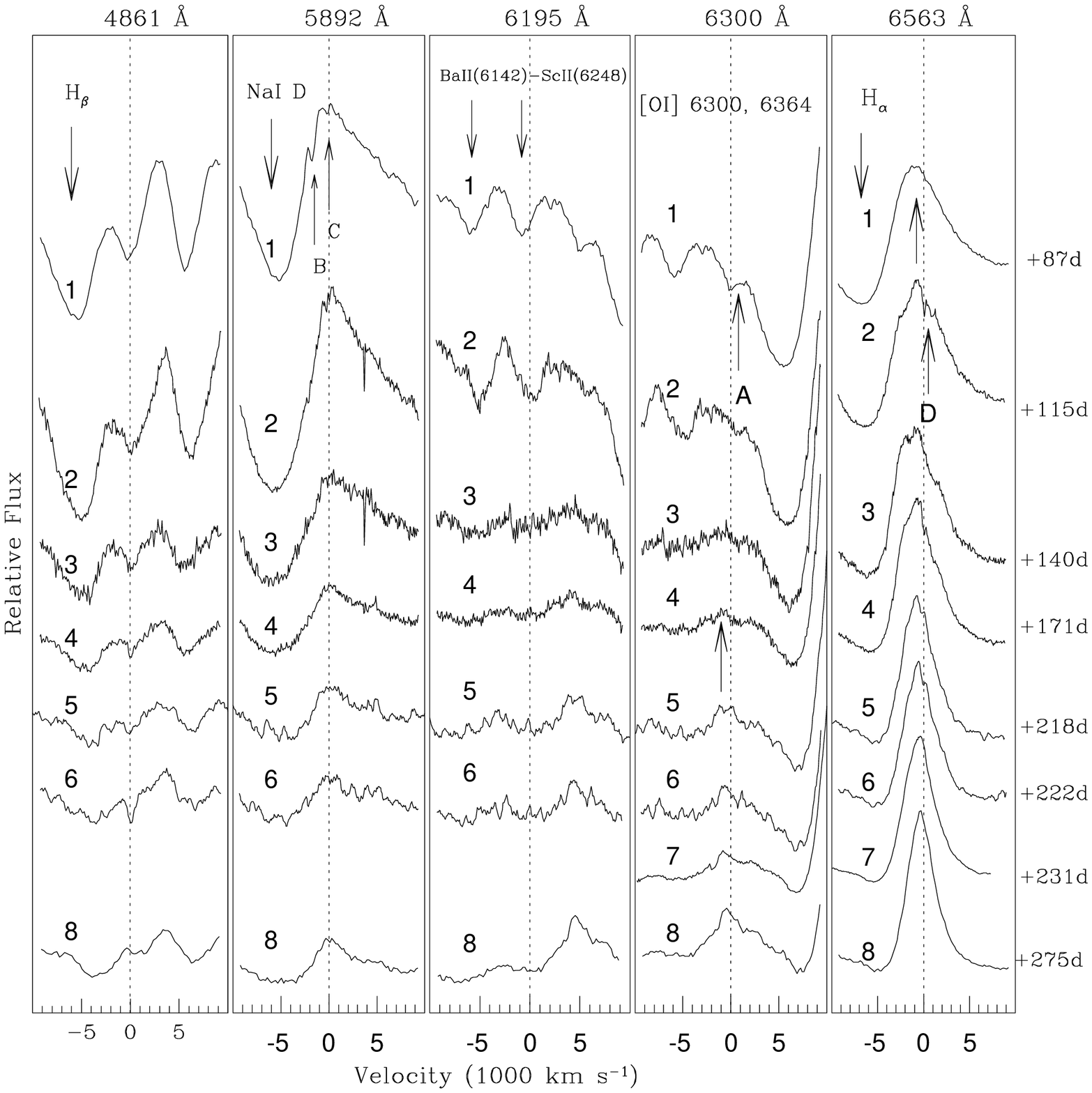}
\caption{Evolution of some important spectral lines of SN 2008gz, during its
  transition from late plateau to nebular phase. The dotted line at zero velocity
  corresponds to the rest wavelength. Upward arrows show emission peaks, while
  downward arrows show absorption dips. The feature 'A' is a high velocity P-Cygni
  component of \ha\,, features 'B' and 'C' are due to \Nai\,D absorption from
  the Milky Way and the host respectively, and the emission feature 'D' is \ha\, emission
  due to underlying \Hii\, region at SN location. The numbering sequences and resolution are :
  1 (+87 d/10\AA), 2 (+115 d/6\AA) 3 (+140d/12\AA), 4 (+171d/6\AA), 5
  (+218d/10\AA), 6 (+222d/10\AA), 7 (+231d/10\AA) and 7 (+275 d/14\AA).}
\label{fig:specevol}
\end{figure*}

 To further illustrate the nature of burst, we show in Fig.~\ref{fig:specevol} (and
 describe below) the velocity profiles of \ha, \hb, \Nai\,D, \Baii\,6142\AA\, 
 and \Oia\,6300, 6364\AA. The absorption dips and emission peaks
 due to SN are marked by downward and upward arrows respectively. 
 For a spherically symmetric burst, the emission peak of P-Cygni profiles should be 
 located at the rest wavelength of the corresponding line, while absorption dip will be 
 blueshifted reflecting the instantaneous velocity of corresponding line emitting region. 
 In our rest frame spectra, \ha\, emission peak is found to be slightly blue shifted by  
 $\sim 406 \kms$ at +87d. Such blue shift in \ha\, emission peak at 
 the early epochs was also observed for other type II SNe 
 (\ie, 1987A \citep{hanuschik87}, 1988A \citep{turatto93}, 1990K \citep{cappellaro95}, 
 1993J \citep{matheson00}, 1998A \citep{pastorello05}, 1999em \citep{elmhamdi03a}, 
 2005cs \citep{pastorello09}, 2007od \citep{andrews10}). It is observed for few 
 low luminosity type II as well, \eg, SN 1999br in their early phases \citep{pastorello04}. 
 On the basis of SN 1987A velocity profile, \citet{chugai88} explained this 
 phenomenon as an effect of diffused reflection of resonance radiation by the 
 expanding photosphere. For SN 1987A at the +85d this velocity was 
 about a few hundred \kms and hence comparable with SN 2008gz. On the other hand for SN
 1998A and SN 1993J \ha\, emission peak velocity at comparable epoch were 
 at least one order of magnitude higher than SN 2008gz. 
 At later epoch (+275d) \ha\, 
 emission peak velocity for SN 2008gz is about $-200 \kms$ which is considerably different 
 from SN 1987A, where redshifted velocity of $+600 \kms$ was observed.

 In most of our \ha\, profile, a narrow emission line (marked by `D' in Fig.~\ref{fig:specevol}) 
 is seen at zero velocity, which is probably due to underlying \Hii\ region and this is consistent 
 with presence of low-luminosity \Hii\, region at SN location revealed by our \ha\, narrow 
 band observation (\S\ref{sec:env}). Such a feature is not seen in well observed
 nearby IIP SNe 2004et \citep{sahu06} and 2004A \citep{gurugubelli08},
 which occurred in the outskirts of their host galaxies.

 \Nai\ D 5892\AA\, P-Cygni nature is prominent in the spectra at all epochs with 
 emission peak located at zero velocity, indicating almost spherical distribution 
 of Na ion in the ejected material. In high S/N spectra (+87d and +115d), two peculiar 
 absorption dip in the SN \Nai\,D emission profile is seen at $-80 \kms$ (marked by B), and 
 at $-1630 \kms$ (marked by C), respectively. These are identified as \Nai\,D absorption 
 respectively due to the interstellar matter in the host galaxy and the Milky Way 
 (\S\ref{sec:dis}). In this regard, we note that due to highly inclined host galaxy 
 ($\Theta_{\rm inc} = 56.2 \degr$), true recession velocity for SN will be different 
 from the adopted ones (1864 \kms), and as measured from \Nai\,D absorption at 
 SN location, it should differ by $\sim100\kms$, however, considering, low dispersion 
 spectra, this will not change any of our conclusions.  
 
 The absorption dips due to an s-process element \Baii\, 6142 \AA\, and for the 
 element \Scii\, 6248\AA\, are clearly visible in +87d and +115d spectra, however,
 they disappeared in the +140d spectrum. This is similar to luminous type IIP SNe 
 2004et \citep{sahu06} where these lines were prominent at +113d and 
 barely observable at +163d and for 1999em \citep{elmhamdi03a} it was visible 
 up to +166d. In contrary to this, low luminosity IIP SNe 1997D, 2005cs and others 
 \citep{turatto98,benetti01,pastorello04}, these features sustain 
 comparatively for longer time and observable till +208d.
 Low luminous type IIP SNe expand with velocity, much
 slower than that of normal type IIP events.
 So, the Ba lines in low luminosity events sustain for longer time just because
 the ejecta takes more time to cool-down. 
 For luminous 
 type II-peculiar SNe 1987A and 1998A, these feature are seen even at later epochs
 beyond +300d. For \sn, probably Ba abundance was low and s-process was not so 
 effective like low luminosity events.
 
 Metastable \Oia\ 6300, 6364 \AA\ lines start to appear at +171d, and its strength 
 increases progressively in spectra at later epochs.
 A normalized profile for +275d 
 is shown in Fig.~\ref{fig:specoii}\footnote{Figure~\ref{fig:specoii} is only 
 available in electronic form.} which is nearly symmetric, indicating a spherically
 symmetric Oxygen ejecta.
 The average FWHM for \Oia\,lines is $\sim$ 2500 \kms.
 Two component Gaussian fit results in ratio of 
 $I(6300)/I(6364) \approx 1.8$, which is at deviation from 
 their strength ratio of 3 expected from transitional probability for a rarefied 
 gas at certain temperature. Smaller ratio for \sn\, may indicate higher opacity 
 for 6300 \AA\ in comparison to 6364\AA.
 We however note that the ratio of $I(6300)/I(6364)$ is not always 3 for type IIP events.
 For the SN 1988A, \citet{spyromilio91}
 showed that at initial epochs when optical depth of the ejecta is very high, value of 
 $I(6300)/I(6364) \approx 0.952$, whereas at late phase, for optically thin ejecta this ratio 
 approaches to 3.03.
 
 Blueshift in \Oia\, lines, particularly at epochs later than +400d (\ie\, in 
 SNe 1999em and 1987A), is interpreted as an effective indicator of dust formation 
 in the SN ejecta due to excessive extinction of redshifted wings of emission 
 lines than the blueshifted ones \citep{lucy91,danziger91}. Between +300-400d, 
 the observed blueshift in oxygen may be due to contamination from the \Feii\,
 mutliplet at 6250\AA.
 The reason of the blue-shift of the oxygen line at early epochs ($\la$ 200 day)
 is still not clear and several hypothesis have been done. In a recent work
 \citet{taubenberger09} described this as a result of residual opacity that remains
 in the inner ejecta. This seems to be the most likely explanation for observed
 blue-shift of oxygen line. Dust formation at 
 an early epoch $\sim$ +300d is also reported for SN 2004et. For \sn\, we estimate 
 a blueshift of $\sim$ 250 \kms, in \Oia\, components at epoch of +275d 
 (see Fig.~\ref{fig:specall} and \ref{fig:specoii}), however due to absence of any 
 other evidence, this is not enough to claim dust formation in the SN 2008gz ejecta.


\section{Photospheric and H-envelope velocities of ejecta} \label{sec:synow}

 We used multi-parametric resonance scattering code {\sc SYNOW} \citep{branch01,branch02,baron05}
 for modelling the spectra of  SN\,2008gz to iterpret spectral features 
 and estimate velocities of layers at different epochs. The algorithm works on  
 the assumptions of spherical symmetry; homologous expansion of 
 \mbox{layers ($v \sim r$);} sharp photosphere producing a black-body spectrum and 
 associated at early stages with a shock wave. In photospheric phase, the spectral lines 
 are formed by the shell above the thick photosphere, but in nebular phase all visible 
 regions are optically thin \citep{branch01}. Each of these two phases of SN evolution 
 can be explained with individual approximations and the modelling of observed spectra 
 needed in different synthetic codes.


\begin{table*}
  \caption{Velocities of photosphere, \ha\, and \hb\, for different epochs of \sn\, evolution. 
   All the parameters are derived from {\sc SYNOW} modelling.} 
  \label{tab:velphot}
  \begin{tabular}{lc cc cc cc}
    \hline
    UT Date &Phase& $v_{\rm ph} = v(\Feii)$& $v_e(\Feii)$& $v(\ha)$&  $v_e(\ha)$&  $v(\hb)$& $v_e(\hb)$  \\
    (yy/mm/dd/) &(days)& \kms&    \kms  &\kms& \kms& \kms& \kms \\ \hline

    2008/11/11& +87& $4200\pm400$          & $1000\pm400$  & $5500\pm200$        & $2100\pm300$        & $4300\pm300$&  $1300\pm100$        \\
    2008/12/08&+115& $3200\pm400$          & $1500\pm500$  & $4700\pm300$        & $1500\pm200$        & $3800\pm600$&  $900^{+300}_{-100}$ \\
    2009/01/03&+140& $4000\pm400$          & $2500\pm1100$ & $5300\pm400$        & $1300\pm300$        & $4000\pm400$&  $1000\pm200$        \\
    2009/02/02&+171& $3500\pm300$          & $1600\pm800$  & $5100\pm200$        & $1300\pm200$        & $3500\pm300$& $1000\pm200$         \\
    2009/03/22&+218& $3100\pm500$          & $1700\pm900$  & $4800\pm400$        & $1500\pm500$        & $3700\pm300$& $1600^{+400}_{-800}$ \\
    2009/03/25&+222& $3100\pm700$          & $1200\pm800$  & $4900^{+700}_{-300}$& $1100^{+300}_{-500}$& $3100\pm700$&  $2000\pm1000$       \\
    2009/04/03&+231& $< 5400$              & ---           & $4800^{+600}_{-200}$& $1600^{+200}_{-600}$& ---         &  ---                 \\
    2009/05/17&+275& $2000^{+200}_{-100}$  & $1500\pm500$  & $3800\pm200$        & $600\pm200$         & $2000\pm100$& $600^{+600}_{-200}$  \\
    \hline
  \end{tabular}
\end{table*}

\setcounter{figure}{8}
\begin{figure}
\centering
\includegraphics[scale=0.42]{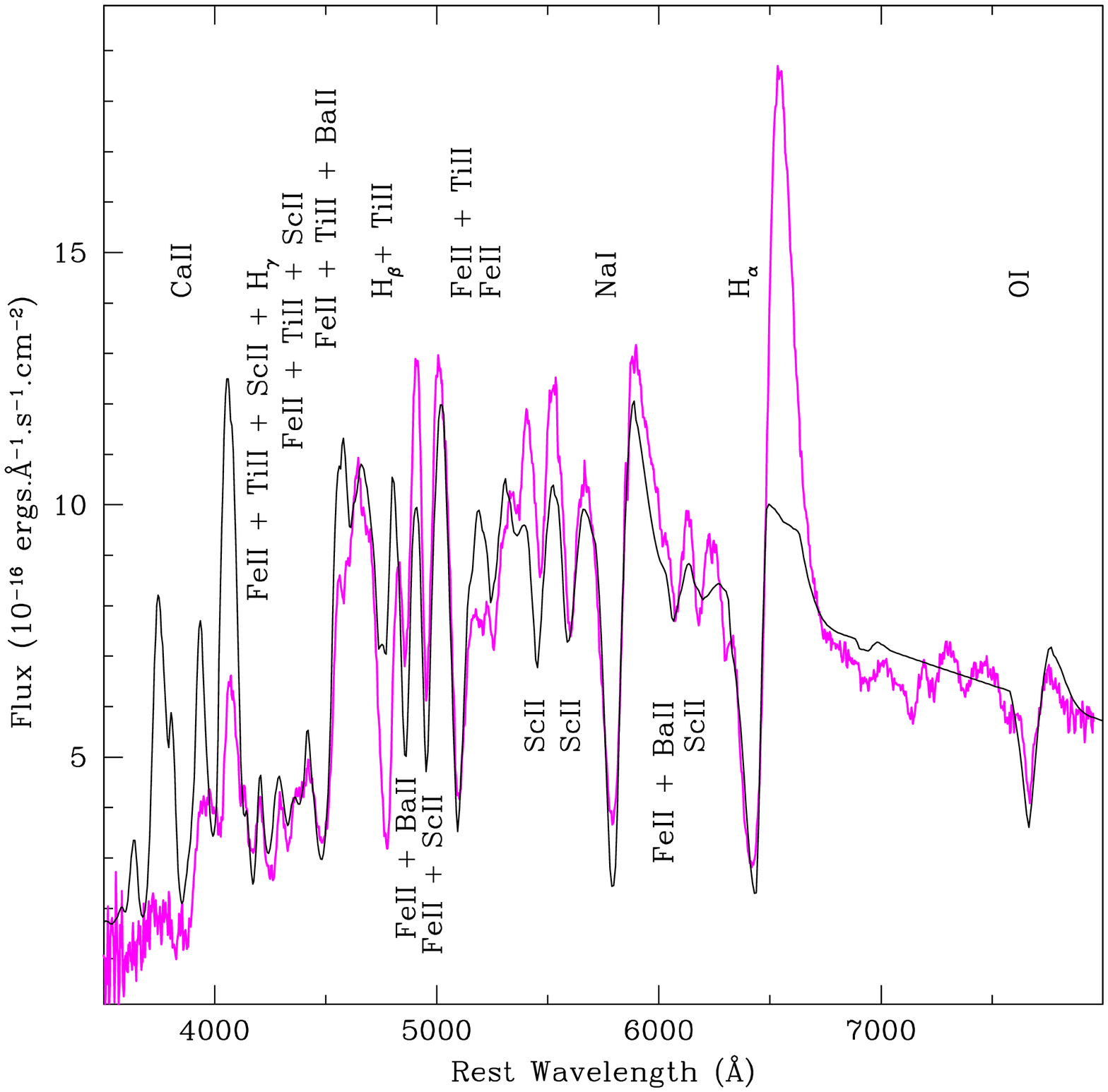}%
\caption{SYNOW modeling of the +87 d spectrum of SN 2008gz. The line
  identification is done after considering the SN as a spherically expanding
  fireball and lines are formed in a region moving ahead of the photosphere.
  Optical depth of individual line is calculated through ``Sobolev
  approximation''.}
\label{fig:synowall}
\end{figure}

\begin{figure}
\centering
\includegraphics[scale = 0.42]{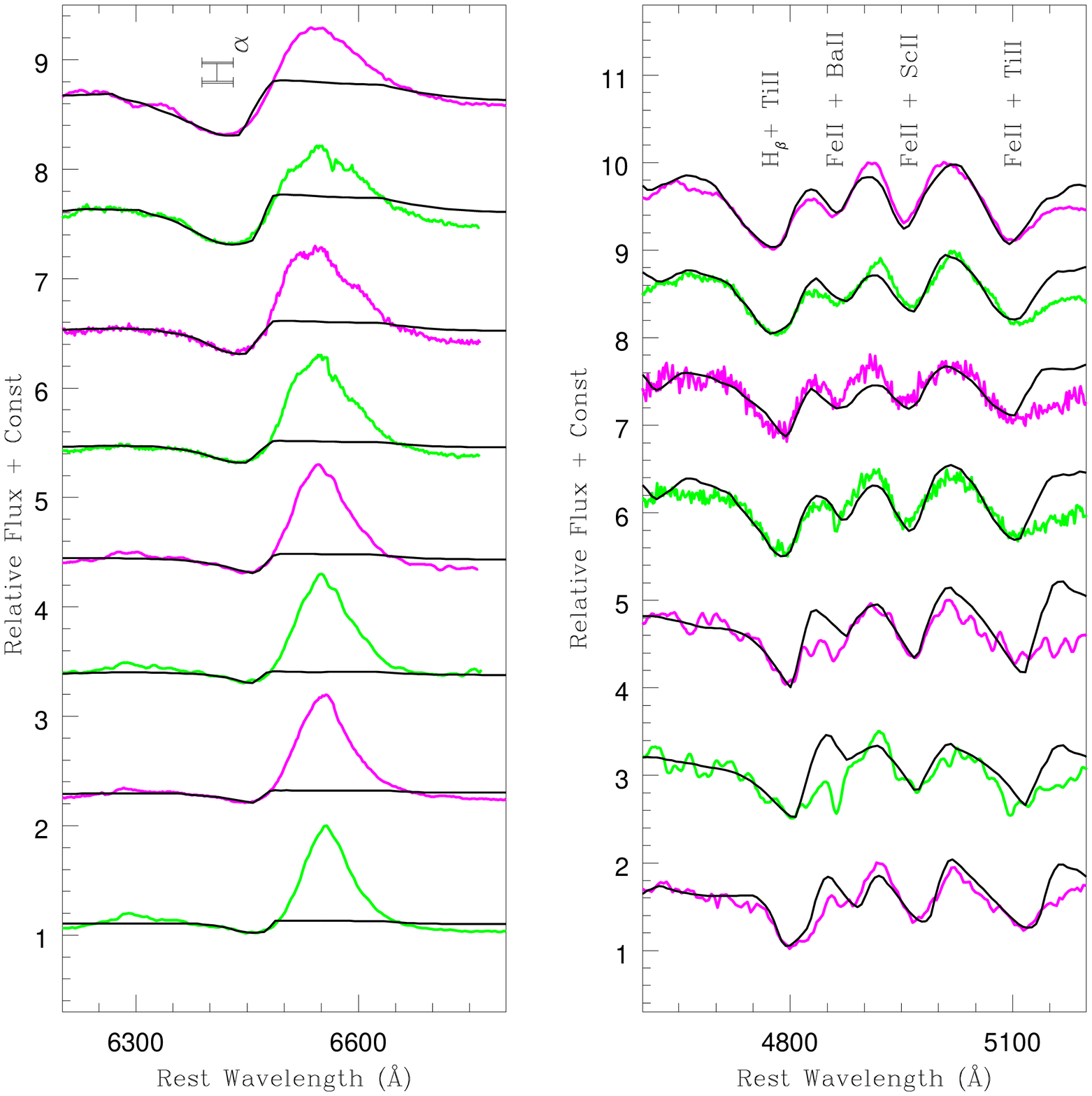}%
\caption{SYNOW models for absorption features of \ha\, (left panel)
  and \hb\, \& \Feii\, (right panel). Blending effect due to \Scii, Tiii and BaII is
  also incorporated in the models. Spectral evolution corresponds (top to bottom) to
  +87, +115, +140, +171
  +218, +222, +231 and +275. \hb\, region is not modeled for +231 d.
 }
\label{fig:synowhi}
\end{figure}

\begin{figure}
\centering
\includegraphics[scale = 0.45]{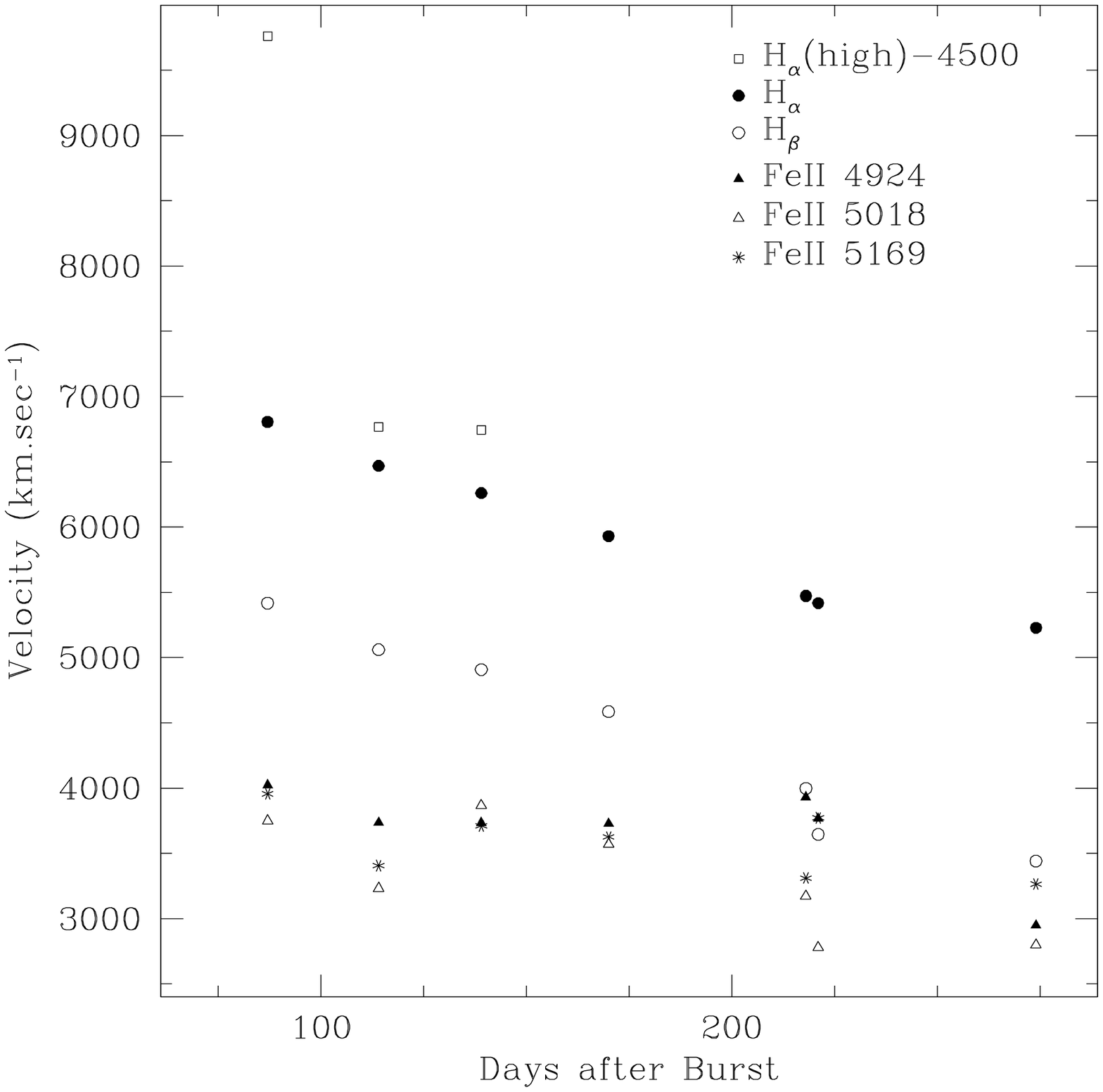}%
\caption{Velocity profiles of different elements in the ejecta of SN 2008gz}
\label{fig:velevol}
\end{figure}


 Our main aim in modeling the spectral features is to estimate the velocities of the layers 
 and that of pseudo-photosphere. It was also noted in Branch, Baron and Jeﬀery (2001) there is 
 no sharp division between the photosphere and the nebular phases. We note the presence of 
 absorption components in Iron and Hydrogen lines at the latest spectra, which can be 
 explained as a result of the decreasing resonance scattering mechanism. Although the 
 resonance scattering codes like SYNOW are not used for describing late time spectra 
 (see for ex. Elmhamdi et al., 2006), we use it to describe only the absorption parts of 
 line profiles. It is not our intention to fit the emission part of the line profiles because 
 this procedure needs to make use of other assumptions and other codes. 

 A preliminary result of SYNOW fit on \sn\, is reported by \cite{moskvitin10}. 
 In Fig.~\ref{fig:synowall}, we present our model fit for +87d spectrum 
 in detached case, \ie, assuming line forming shells of ionised gases moving ahead of
 the photosphere. Most of the spectral features (particularly absorption minima and the 
 continuum) are produced well. All the identified spectral features are same as marked 
 in Fig.~\ref{fig:specid}. We tried out undetached cases \citep{sonbas08} as well 
 and also attempted changing density laws (exponential and power) and we found that it 
 had very little effect while fitting the absorption minima.
 In order to obtain precise velocity measurements of hydrogen layers, we modeled the profiles 
 of \ha, \hb\, and \Feii\, independently (see Fig.~\ref{fig:synowhi}) following  
 $\tau \sim exp(-v(r)/v_e)$, where $\tau$ is optical depth and $v_e$ is the e-fold 
 velocity. We also incorporated \Tiii, \Scii, and \Baii\, ions to model multi-minima
 absorption features around \hb. 
 For \Feii\, lines, we noticed that velocities of various absorption 
 features are similar or have insignificant differences and hence we used averaged \Feii\, 
 values to represent the photospheric velocity ($v_{\rm ph}$) \citep{branch01,elmhamdi06}. 
 Estimations of photospheric and envelopes velocities of different layers are given in 
 Table~\ref{tab:velphot}. Uncertainties in the estimates take into account the noise 
 in the spectra. 
 
 We also estimated photospheric and H-envelope velocities using IRAF by directly locating 
 the absorption minima and the same is shown in Fig.~\ref{fig:velevol}. The velocity  
 for \Feii\, 4924, 5018 and 5169 lines, range from $\sim$ 4000 \kms at +87d to around 
 3000 \kms at +275d. These values are similar to the values estimated above 
 from {\sc SYNOW} modelling. For \hb\, and \ha\, layers our values are consistently 
 higher by about 1500 \kms at all epochs than that derived from {\sc SYNOW}.
 This discrepancy may arise due to contamination of true absorption minima by the 
 emission component of P-Cygni profile of \ha\, and \hb\, and hence it is likely that the true
 blueshift would be overestimated while using absorption minima. 

 For \sn\, the \ha\, velocity at +87d is $\sim$ 6800 \kms, while for SNe 2004et, it
 is $\sim$ 6000 \kms \citep{sahu06}. \hb\, also shows higher expansion velocity
 at comparable epochs. Similarly the photospheric velocity at day +50 is $\sim$ 3700 \kms, 
 while it is 4000 \kms at +87d for \sn. So, even by considering an overestimate of the plateau 
 period about 25 days (\ie\, for $t_{i} \sim 95$d), the photspheric and H-envelope velocity for \sn\ seem 
 to have comparable or higher values than SN 2004et. The photospheric velocity is a good indicator 
 of the explosion energy \citep[see][]{dessart10} and hence \sn\, has explosion 
 energy similar to that of SN 2004et, $\sim 2.3\times10^{51}$ erg \citep{utrobin09} or higher. 
 \cite{utrobin07} obtains an explosion energy of $\sim 1.3\times 10^{51}$ erg for SN 1999em, 
 which has comparatively lower expansion velocity than to SN 2004et at similar epochs.


\section{Distance and extinction of SN 2008gz} \label{sec:dis}

\setcounter{figure}{12}
\begin{figure}
\centering
\includegraphics[scale = 0.42]{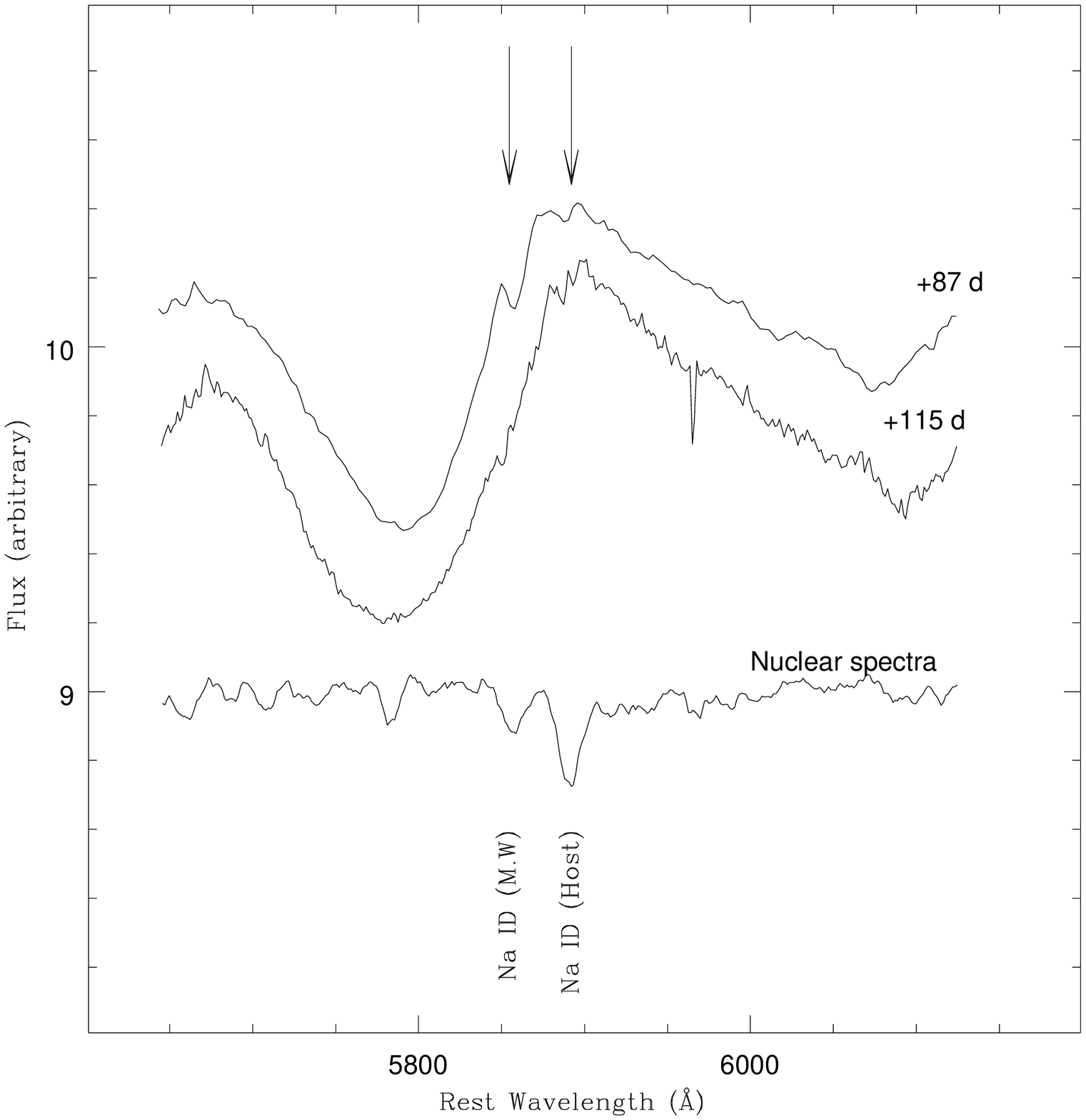}%
\caption{Rest wavelength spectra of \sn\, (+87 d/10\AA\, and +115 d /6\AA\,) and the centre of
         host galaxy (+170 d /6\AA\,). \Nai\,D absorption due to interstellar matter
         of host galaxy ($\sim$ 5892\AA) and the Milky Way ($\sim$ 5854\AA) is indicated.}
\label{fig:specred}
\end{figure}
 Spectrum of the nuclear region of the host galaxy taken 
 on +170d (see Fig.~\ref{fig:specgal}\footnote{Fig.~\ref{fig:specgal} is available only 
 in electronic form.}) was used to estimate $cz_{\rm helio}$, the heliocentric velocity.
 Employing five nebular emission lines and prominent absorption features and using 
 the 5577 \AA\, skyline as a reference wavelength, we obtain $cz_{\rm helio}$ of $1891\pm14 \kms$ 
 This is in agreement with 18 other radio and optical measurements of $cz_{\rm helio}$ of NGC 3672 
 (in range of 1400 to 2000 \kms) listed in HyperLEDA\footnote{http://leda.univ-lyon1.fr/}. 
 The Combined measurement gives a mean $cz_{\rm helio}$ of $\sim1864\pm19$ \kms and 
 it corresponds to a corrected (Local Group infall into Virgo) distance 
 of $\sim25.65\pm2.93$ Mpc\footnote{The cosmological 
 model with H$_0$ = 70 km.s$^{-1}$.Mpc$^{-1}$,$\Omega_{m}$ = 0.3 and 
 $\Omega_{\Lambda}$ = 0.7 is assumed throughout the paper and the uncertainty corresponds to
 a local cosmic thermal velocity of 208 \kms \citep{terry02}.}.
 NED\footnote{http://nedwww.ipac.caltech.edu/} lists four distance measurements
 based on \Hi\, Tully-Fisher relation, with a mean of $25.25\pm4.0$ Mpc, which is in agreement
 with the above kinematic estimate, we therefore, adopt uncertainty
 weighted distance of $25.5\pm2.4$ Mpc for \sn.

\begin{table}
  \caption{Equivalent width measurement of \Nai\,D absorptions in the spectra of \sn\, and
   the host galaxy. Last row provides the uncertainty weighted EW of \Nai\,D absorption 
   in the direction of \sn\, due to Milky Way and the host.} 
  \label{tab:nadew}
  \begin{tabular}{lc cc}
    \hline
    UT Date &  Phase& EW (MW)& EW(host) \\
    (yy/mm/dd/)& (days)& \AA& \AA \\ \hline

    2008/11/11& +87& $1.39\pm0.34$& $0.23\pm0.28$ \\
    2008/12/08&+115& $1.21\pm0.69$& $0.28\pm0.50$ \\
    2009/02/02&+171& $1.32\pm1.29$& --            \\ \\
    \hline
    Weighted EW          &    & $1.29\pm0.29$& $0.23\pm0.24$ \\
    \hline
  \end{tabular}
\end{table}

 The Galactic reddening in the direction of \sn\, as derived from the 100 \mum\, all sky 
 dust extinction map of \citet{schlegel98} is estimated as \ebv\, = $0.041\pm0.004$ mag.
 Additionally, we could also determine reddening in the direction of \sn\, from equivalent widths 
 of \Nai\,D absorption lines present in the spectra of \sn\, (+87d and +115d) and
 the centre of the host galaxy (see Fig.~\ref{fig:specred}). D$_1$ (5889.95\AA) and 
 D$_2$(5895.92\AA) component of \Nai\,D is not resolved in +87d spectra and it is seen 
 as narrow absorption features overlaid on the broad P-Cygni emission wings of \Nai\,D due 
 to SN. In rest wavelength plot the host galaxy contribution is seen at $\sim -80 \kms$, 
 while the Galactic contribution is at $-1630 \kms$. In +115d spectrum, both components of
 \Nai\,D are resolved. Intriguingly, the Galactic component appear to split into two --
 a stronger component at $\sim -1896 \kms$ (due to Milky Way ISM) and a weaker component
 at $\sim -1021 \kms$, possibly due to inter-galactic medium. In the +170d spectra, 
 the Galactic \Nai\,D absorption appear as a single component.
 Estimated total \Nai\,D equivalent widths (EW) are reported in Table~\ref{tab:nadew}.
 Quoted errors in EW are photon-noise dominated 
 RMS uncertainty derived following \citet[see their Eq. 6]{vollmann06}. It is
 seen that the EW contribution due to host ($0.23\pm0.24$\AA) is smaller in comparison 
 with the total Galactic contribution ($1.29\pm0.29$\AA). 

 It is known that the EWs of interstellar absorption bands is well correlated with 
 the reddening \ebv\, estimated from the tail of SNIa colour curves \citep{barbon90,richmond94}
 and by employing empirical relations established 
 by \citet{turatto03}, $\ebv = -0.01+0.16{\rm EW}$ (where EW is in \AA)\footnote{In
 \citet{turatto03} there are two relations $-$ one is with low slope and other with a
 high slope. In this work we have
 considered the lower slope, because it is well sampled and matches with the previous works in
 this direction \citep{barbon90}.}, we obtain 
 Galactic \ebv\, contribution as $0.20\pm0.05$ mag and host galaxy as $0.03\pm0.04$ mag. 
 The Galactic \ebv\, derived in this way is larger than that derived
 from Schlegel map. Considering the normal extinction law (R$_V =$ 3.1) and Schlegel value
 of \ebv\, $=$ 0.041 for the Milky Way; the EW (host/Galactic) ratio would suggest a slightly
 lower reddening in the host (\ebv\,$\leq$ 0.01 mag). However the host galaxy value of 0.03 mag
 may not be ruled out in case of a different dust to gas ratio for the host.   

 For this work, we will adopt a conservative value 
 of \ebv=$0.07\pm0.04$ mag, obtained by adding Galactic (Schlegel) and host galaxy (\Nai\,D) 
 contribution. This corresponds to visual extinction ($A_V$) of $0.21\pm0.12$ by assuming ratio of 
 total-to-selective extinction $R_V=3.1$ \citep{cardelli89}.


\section{Temporal Evolution of colour and bolometric luminosity} \label{sec:bol}
\begin{figure}
\centering
\includegraphics[scale = 0.45]{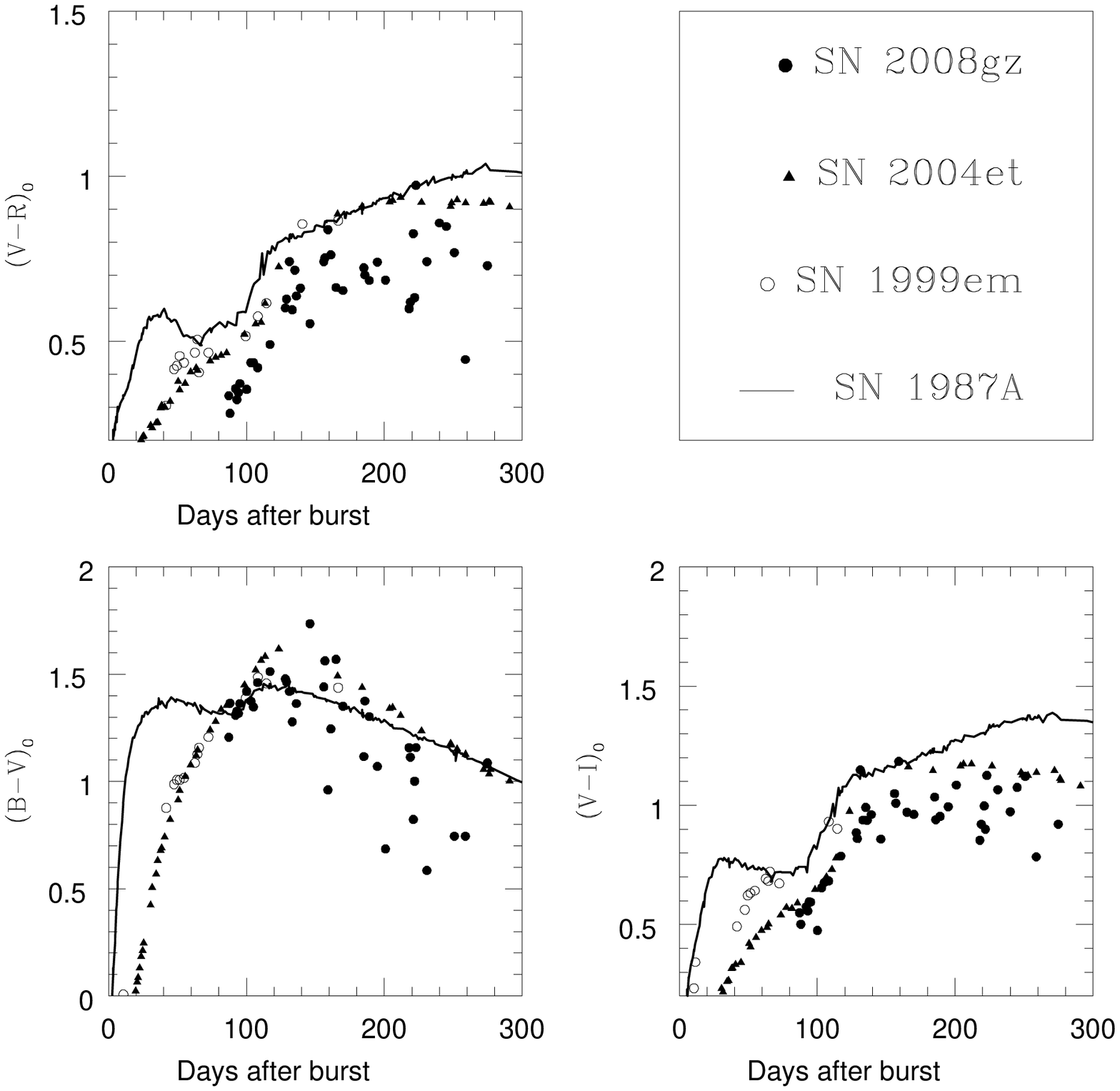}
\caption{Temporal variation of colour of SN 2008gz. Also shown are the other
 core-collapse supernovae, SN 1987A, SN 1999em and SN 2004et.}
\label{fig:colcur}
\end{figure}
\begin{figure}
\includegraphics[scale = 0.43]{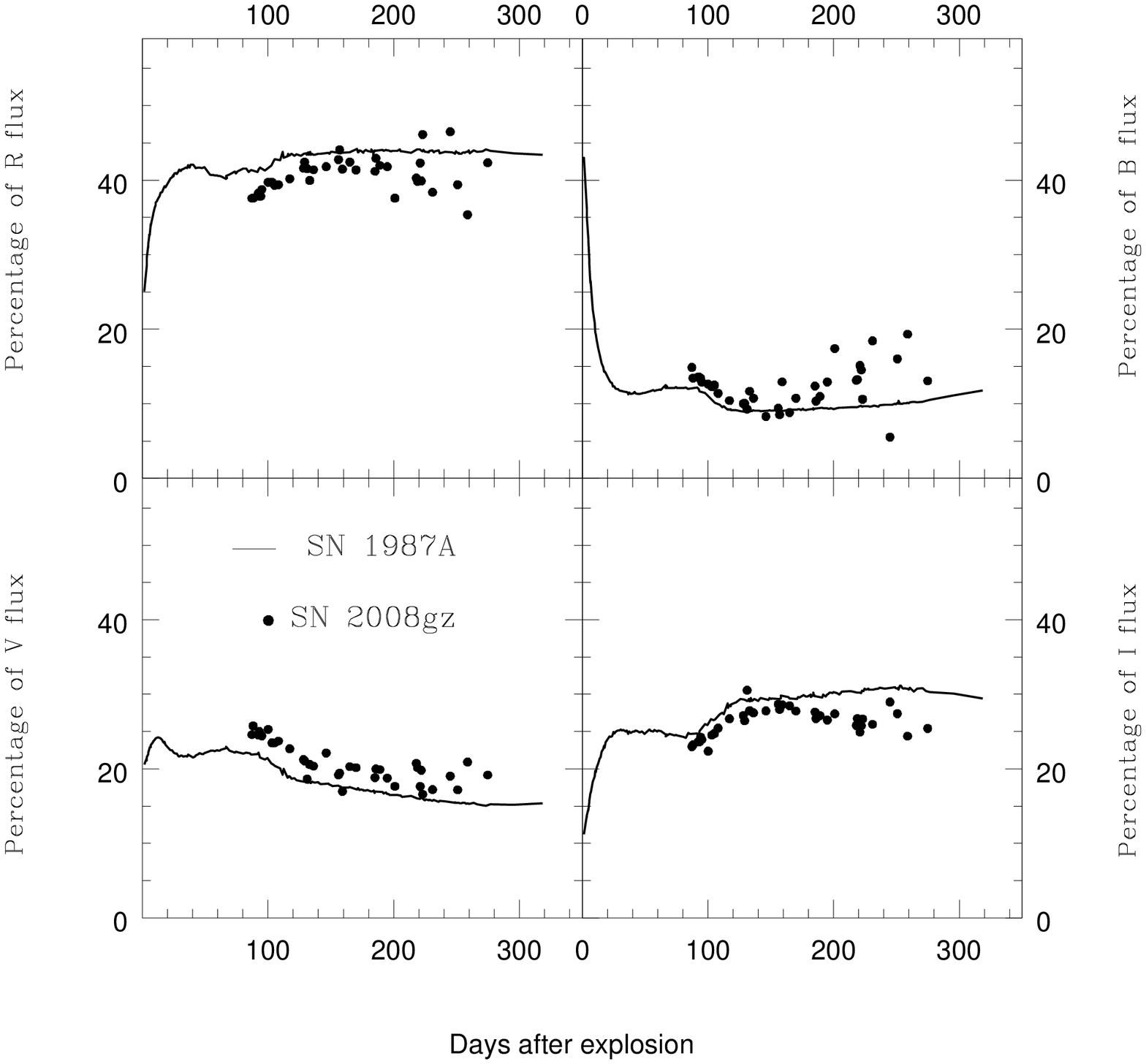}
\caption{Flux contribution in percentage, in $BVRI$ bands of SN 2008gz
  along with a comparison to SN 1987A}
\label{fig:lightfcur}
\end{figure}
\begin{figure}
\centering
\includegraphics[scale = 0.44]{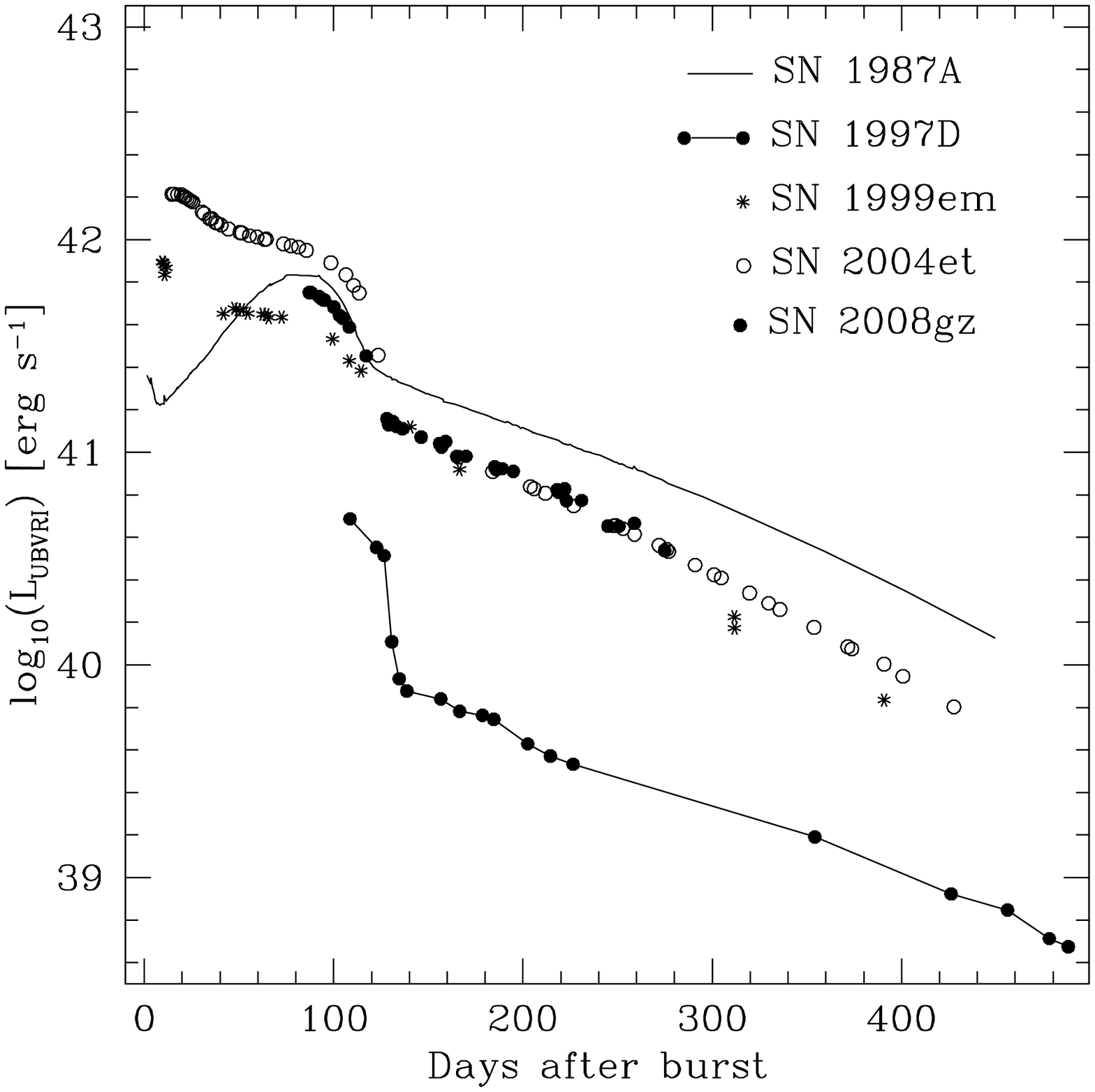}
\caption{Bolometric light curve of SN 2008gz. For comparison, we also show
   light curves of SN 1987A, SN 1997D, SN 1999em and SN 2004et.}
\label{fig:bolcur}
\end{figure}
 Fig.~\ref{fig:colcur} shows reddening corrected colour evolution of SN 2008gz.
 For comparison, we also show reddening corrected colour curve of 
 SN 1987A \citep{suntzeff90}, SN 1999em \citep{elmhamdi03a} and SN 2004et \citep{sahu06}, 
 for $E(B-V)$ of 0.15, 0.10, and 0.41 mag and explosion epochs of JD 2446849.82, 
 JD 2451480.5 and JD 2453275.5 respectively. Though the colour curves of SN 2008gz have 
 large scatter but overall 
 nature of its temporal evolution is prominent. SN 2008gz follows general
 trend of colour evolution i.e. a steep and rapid decrease from blue (high temperature)
 to red (low temperatures) colours similar to SNe 2004et and 1999em. 
 $(B-V)_{0}$ becomes redder from 1.2 mag at +87d to about 1.7 mag 
 at +140d and it follows trend similar to SNe 2004et and 1999em. The overall trend in colour 
 evolution of SN 2008gz between end of plateau and middle of the tail is similar
 to that of peculiar type II SN 1987A. In nebular phase, $(B-V)_{0}$ turns blue rather 
 rapidly, $-1\,{\rm mag}\,(100{\rm d})^{-1}$ and this arises due to suspected flattening 
 in $B$ light and a shallow decay in $V$, $R$ and $I$ bands. The $(V-R)_{0}$ colour is found 
 to be consistently bluer than for other IIP SNe. For $(V-R)_{0}$ and $(V-I)_{0}$, the 
 increment is quite shallow during the transition from plateau to nebular phase. This is 
 in contrary to the colour evolution of low luminosity type II-peculiar SNe 1997D and 1999eu, 
 where a steep rise and excess in colour has been noticed \citep{pastorello04,misra07}. 

 Bolometric luminosity is essential to estimate total optical radiant energy in the 
 explosion and also at the tail phase it is a good estimator of radioactive 
 \nickel\, synthesized in the explosion. To a good approximation, the integration of 
 extinction corrected flux in $UBVRI$ at a given epoch gives a meaningful estimate 
 of bolometric luminosity. The extinction corrected $BVRI$ magnitudes were converted into 
 fluxes using zeropoints
 given by \citet{bessell98} and the total $BVRI$ flux is obtained by interpolating and 
 integrating fluxes between 0.4 to 0.85 \mum. Fig.~\ref{fig:lightfcur} shows percentage 
 flux contribution in different passbands and the overall trend is found to be similar 
 to that of SN 1987A. From plateau (+87d) to nebular (+140d) phase, the flux 
 contribution declines from $\sim$ 15 to 10\% at $B$, from $\sim$ 25 to 20\% at $V$, while 
 it ascends from $\sim$ 36 to 42\% at $R$, from $\sim$ 22 to 30\% at $I$. In nebular phase 
 (until +270d), the flux contributions remain constant. For nearby SNe 1987A and 2004et, 
 large ($\sim$ 40\%) flux contribution in $U$ and $B$ bands are observed 
 at initial epochs, which reduces to a few percent by +60d. Though, towrads later epochs, 
 when SN ejecta becomes optically thin, a little enhancement (about 5\%) in $U$ and $B$ bands is 
 also noticeable \citep{misra07}. Therefore, for SN 2008gz, we have constructed $UBVRI$ 
 bolometric light curve after making a constant (5\%) contribution from $U$ band over the 
 period of our observation. No correction for flux contributions in the ultraviolet and 
 near-infrared region were made as they become significant respectively in early and late 
 phases of the light curve evolution.

 Fig.~\ref{fig:bolcur} shows the $UBVRI$ bolometric nature of SN 2008gz along with other 
 type II events. Different behavior of these events is clearly evident and it provides 
 constraint on the synthesized radioactive \nickel\, as well as explosion energy of SNe.
 Tail luminosity of \sn\, is similar to SN 2004et, while the plateau luminosity is about
 0.2 dex fainter. Explosion parameters for \sn\, are estimated in next sections.


\section {Physical Parameters} \label{sec:par}

\subsection {Amount of ejected radioactive Nickel} \label{sec:nick}

 The nebular phase light curve of type II SNe is mainly governed by the radioactive 
 decay of \nickel\, to \cobalt\, to \iron\, having half-life of 6.1 and 111.26 days 
 respectively and hence the tail luminosity is directly proportional to the amount 
 of \nickel\, synthesized by explosive burning of Si and O during shock 
 breakout \citep{arnett80,arnett96}.

 By using tail luminosity, \nickel\, mass 
 can be derived following the method described by \citet{hamuy03} applied under the 
 assumption that all the $\gamma$-rays emitted during the radioactive decay make the ejecta 
 thermalised. For \sn\, using the $V$ band magnitude at +200d, corrected for extinction 
 ($A_V = 0.21\pm0.12$ mag; \S\ref{sec:dis}), a bolometric correction of $0.26 \pm 0.06$ mag
 \citep{hamuy01} during nebular phase, and a distance modulus of $32.03\pm0.21$, we 
 derive tail luminosity of $1.51\pm0.29\times10^{41}\,{\rm erg\,s^{-1}}$ and this, for the 
 plateau duration of 115 days, results in Ni mass $M_{\rm Ni} =0.067\pm0.012$\msun. 

 We can estimate the mass of \nickel\, by a direct comparison with that of SN 1987A for which
 it is accurately determined as 0.075\msun. We note that the temporal evolution of 
 flux contribution in different bands for SN 2008gz is comparable with that for SN 1987A
 (\S\ref{sec:bol}) and consequently we can safely assume that at a comparable epoch 
 the ratio of their luminosities is equal to the ratio of synthesized \nickel\, mass. 
 SN 2008gz attains the deep nebular phase beyond 150 days after the burst. The last
 observation was also performed nearly at +275 day. Hence, the mean ratio between
 tail $UBVRI$ bolometric luminosity of SN 2008gz ($\sim$ 150$-$275) and that of SN 1987A
 is about 0.539. This implies that for SN 2008gz ejected \nickel\, mass
 [0.539$\times$0.075] $\approx$ 0.041\msun.

 By taking a sample of ten IIP SNe, \citet{elmhamdi03b} show that the steepness of V-band light 
 curve slope (defined as S=d$m_{\rm V}$/dt) at the inflection time ($t_{i}$) is anti-correlated 
 with \nickel\, mass (${\rm log} M_{\rm Ni} = -6.2295 {\rm S} -0.8147$). For \sn\, we 
 have well sampled transition phase and get a value of ${\rm S}=0.075\pm0.036$, mag d$^{-1}$
 (see Fig.~\ref{fig:steep}\footnote{Fig.~\ref{fig:steep} is only available in electronic form.}) 
 which corresponds to $M_{Ni}=0.052\pm0.01$\msun. Considering the uncertain values of extinction
 towards \sn\,, we note that for SNe 2004A and 2003gd, \citet{hendry06} finds that
 the \citet{elmhamdi03b} scheme gives somewhat lower values. Further, based on plateau luminosity, 
 a linear correlation, ${\rm log} M_{\rm Ni} = -0.438 M_{V}(t_{i}-35)-8.46$ found by 
 \citet{elmhamdi03b} provides a value of \nickel\, mass as 0.051\msun\,for $M_{V}$ of 
 $-16.37\pm0.24$ at +87d ($t_{i}-28$). 
 Taking average of above four estimates, we get the amount of produced \nickel\, mass
 $=$ $0.05\pm0.01$\msun\,. 
 For estimation of other physical quantities we assume that above amount of \nickel\, was produced
 by \sn.
 Though we would like to state that adopting smaller plateau duration (\S\ref{sec:phot.curve})
 will reduce the amount of ejected radioactive  $^{56}$Ni which will further propagate in
 determination of progenitor properties (\S\ref{sec:proj}).

\subsection {Environment of the progenitor} \label{sec:env}
\setcounter{figure}{17}
\begin{figure*}
\centering
\includegraphics[scale=1]{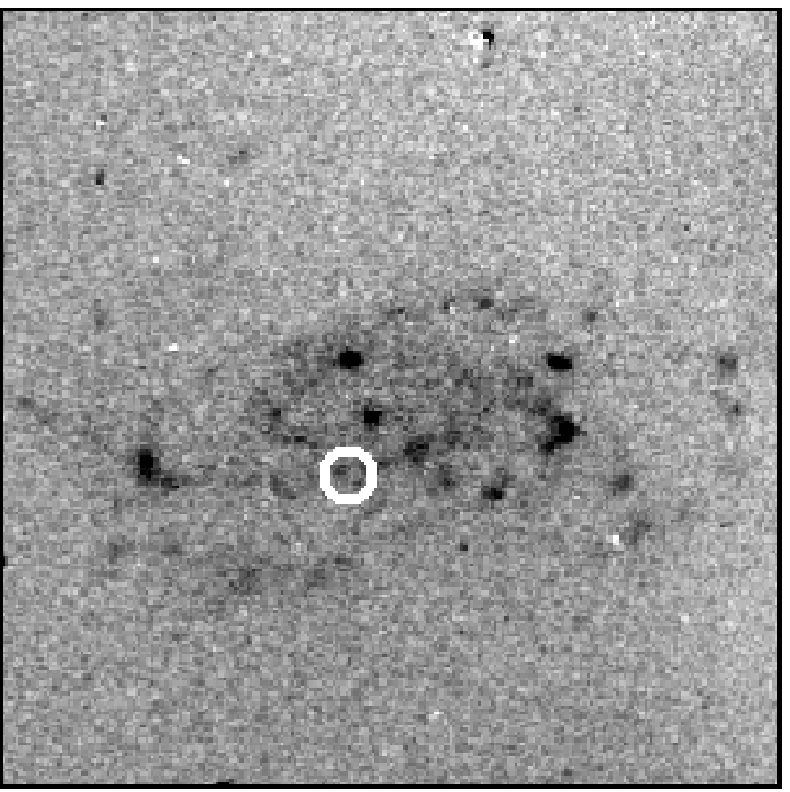}
\includegraphics[scale=.73]{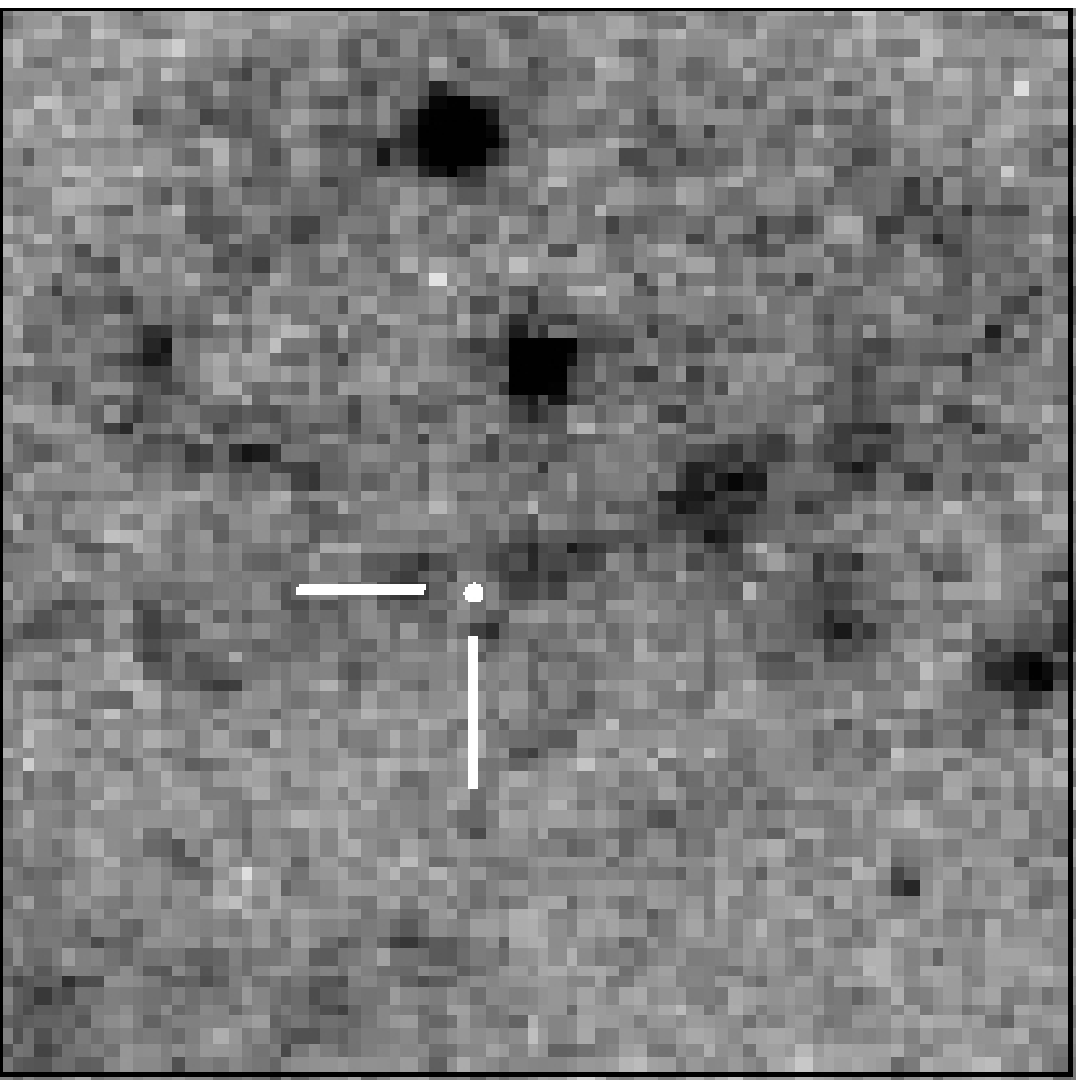}
\caption{Continuum subtracted H$\alpha$ image of NGC 3672. In left panel ($\sim 3\arcmin\times3\arcmin$)
  five bright \Hii\, regions are visible. The SN position is marked with a circle. In right panel
  ($\sim 1\arcmin\times1\arcmin$) zoomed in image of SN is shown. SN location is marked with a dot.}
\label{fig:halpha}
\end{figure*}
 Constraints on the nature of progenitors of core-collapse SNe are derived from the studies of 
 environments in which they occur. For example, by correlating the position of explosion sites
 with that of the sites of recent star formation as traced by \ha\ line emission, it is
 found \citep[see e.g.][]{james06,kelly08,anderson08} that core-collapse events are excellent 
 tracer of star formation, and type Ib/c are more likely to be associated with regions
 of high surface brightness or high \ha\ emission than type II SNe. \citet{anderson09}
 found that type IIP events are likely to be more centrally concentrated than other II-sub types. 
 \sn\, occurred in spiral arms of the host galaxy at a deprojected galactocentric distance of 
 2.8 kpc (within half light radius) and the oxygen abundance ([O/H]=12+log$N_{\rm O}/N_{\rm H}$) 
 of the galactic ISM at the position of SN is estimated as 8.6 (derived from the O/H-$M_{\rm B}$
 relation proposed by \citet{pilyugin04}), which is close to the solar abundance
 [O/H] of 8.65 \citep{asplund09}. 

 In order to further investigate the level of \ha\, emission level, we observed \gal\, 
 in narrow-band \ha-line ($\lambda_{c}$ = 6551\AA) and \ha-red ($\lambda_{c}$ = 6650\AA) having 
 FWHM of 83\AA\, and 79\AA\, respectively. We used 2k CCD camera mounted with 1m ST, 
 Nainital on 18 February 2010 (+560 d). A total exposure of 1 hr in each \ha-line and \ha-red 
 filters were taken along with several bias and sky flats. Raw images were corrected for bias
 and flats using IRAF. FWHM of stellar PSF (seeing) varied from 1\farcs8 to 2\farcs2. Images
 were combined to improve signal-to-noise ratio and the continuum subtraction was done using ISIS. 
 Our narrow-band filter-set were not customized for extra-galactic work, and we expected \ha-red to 
 contain emission line fluxes, as at the redshifted wavelength 
 ($\sim41$\AA\, at \ha\,) of \gal\,, the \ha-red filter had a transmission of 25\%, 40\% and 80\% 
 respectively for \Nii\,6548\AA\,, \ha\, and \Nii\,6589\AA. We could verify this by 
 subtracting \ha-line frame from the broad-band $R$ and $V$ frames of SN taken 
 on 14 February 2010, which gave no residual, while \ha-red frame showed residuals. 
 Fig.~\ref{fig:halpha} shows the continuum subtracted image of \sn\, showing 
 contributions from \ha\,+\Nii\, emissions. A close up view of SN location is also shown. The sky
 is at level of 0 while the peak flux is around 22 counts. Five prominent regions of \ha\, emission 
 having peak counts above 15 are clearly apparent. \sn\, position is at a level of
 8 counts, and hence it belongs to a low-luminosity \Hii\, regions. This is also evident from 
 the early epoch spectra, in which an emission peak (not so prominent) of zero velocity is 
 seen.

\subsection {Properties of progenitor star} \label{sec:proj}

 Accurate estimates of explosion parameters require detailed hydrodynamical 
 modeling of the optical light curves, though the analytical relations (based on a few well 
 modeled IIP events) correlating the physical parameters explosion energy, pre-SN 
 radius and total ejected mass on the one hand and the observable 
 quantities, plateau duration, mid-plateau V-band magnitude ($(M_V)_{\rm mp}$) and 
 mid-plateau photospheric velocity ($v_{\rm mp}$) on the other hand are proposed to exit 
 \citep[see e.g.][]{popov93,litvinova85,nadyozhin03}. For \sn\,, we can only provide
 an approximation of these observables. The estimates of $(M_V)_{\rm mp}$ and $v_{\rm mp}$
 can be made by using the \nickel\, mass estimate and by employing following empirical relations 
 derived (by minimizing least squares) from the data given 
 in \citet[][see their Fig. 3 and 4]{hamuy03}.
  
  \[{\rm log} (M_{\rm Ni}) = -0.385 \times (M_V)_{\rm mp} - 7.749\] 
  \[{\rm log} (M_{\rm Ni}) = 2.771 \times {\rm log} (v_{\rm mp}) - 11.425\]

 For $M_{\rm Ni}$ $=$ 0.05 $\pm$ 0.01 M$_\odot$, we derive 
 ($M_V)_{\rm mp}$ $=$ ${-16.7}^{-0.2}_{+0.3}$ mag and 
 $v_{\rm mp}$ $=$ {4503}$^{+306}_{-348}$ km s$^{-1}$. 
 This value of absolute magnitude is consistent with those obtained from
 the first photometric V point (\S\ref{sec:comp}).
 These estimates, along with a plateau duration of 115 days
 provides \citep{litvinova85} burst energy $\sim$ ${2.5}^{+0.8}_{-0.7}\times10^{51}$ erg, 
 ejected mass $\sim {34}^{+10}_{-8}$ M$_\odot$ and pre-SN radius
 $\sim {167}^{+106}_{-61}$ R$_\odot$.
 The explosion energy derived in this way is consistent with the one expected
 from photospheric velocity (see \S\ref{sec:synow}), however, the ejecta mass
 is larger in comparison to the typical progenitor mass range (8.5-16.5\msun)
 estimate derived from pre-explosion SN images \citep{smartt09a} and this
 may arise due to uncertainty in above measured parameters.
 
 \cite{dessart10} demonstrate that by employing the explosion energy estimate,  
 the observed line width of \Oia\, and artificially generated radiation hydrodynamic 
 simulations of core-collapse SNe, it is possible to put an upper limit on the 
 main-sequence mass of the progenitor. For \sn, the Oxygen ejecta velocity 
 of $\sim 1250 \kms$ (HWHM of \Oia\, profile, see \S\ref{sec:lines}) 
 (which is slightly higher than $\sim 1000\kms$ observed for 
 SNe 2004et and 1999em at around +330 d) and assuming an explosion energy 
 of $3\times10^{51}$ erg s$^{-1}$ provide a main-sequence mass of
 15 (12) respectively for non-rotating (rotating) pre-SN models while 
 assuming an ejecta velocity of 1500\kms for \Oia\, gives an 
 upper limit of 17(13) \msun, which is consistent with that derived
 using pre-explosion images \citep{smartt09a}.


\section{Comparison with other core collapse supernovae} \label{sec:comp}
\begin{figure}
\centering
\includegraphics[scale = 0.46]{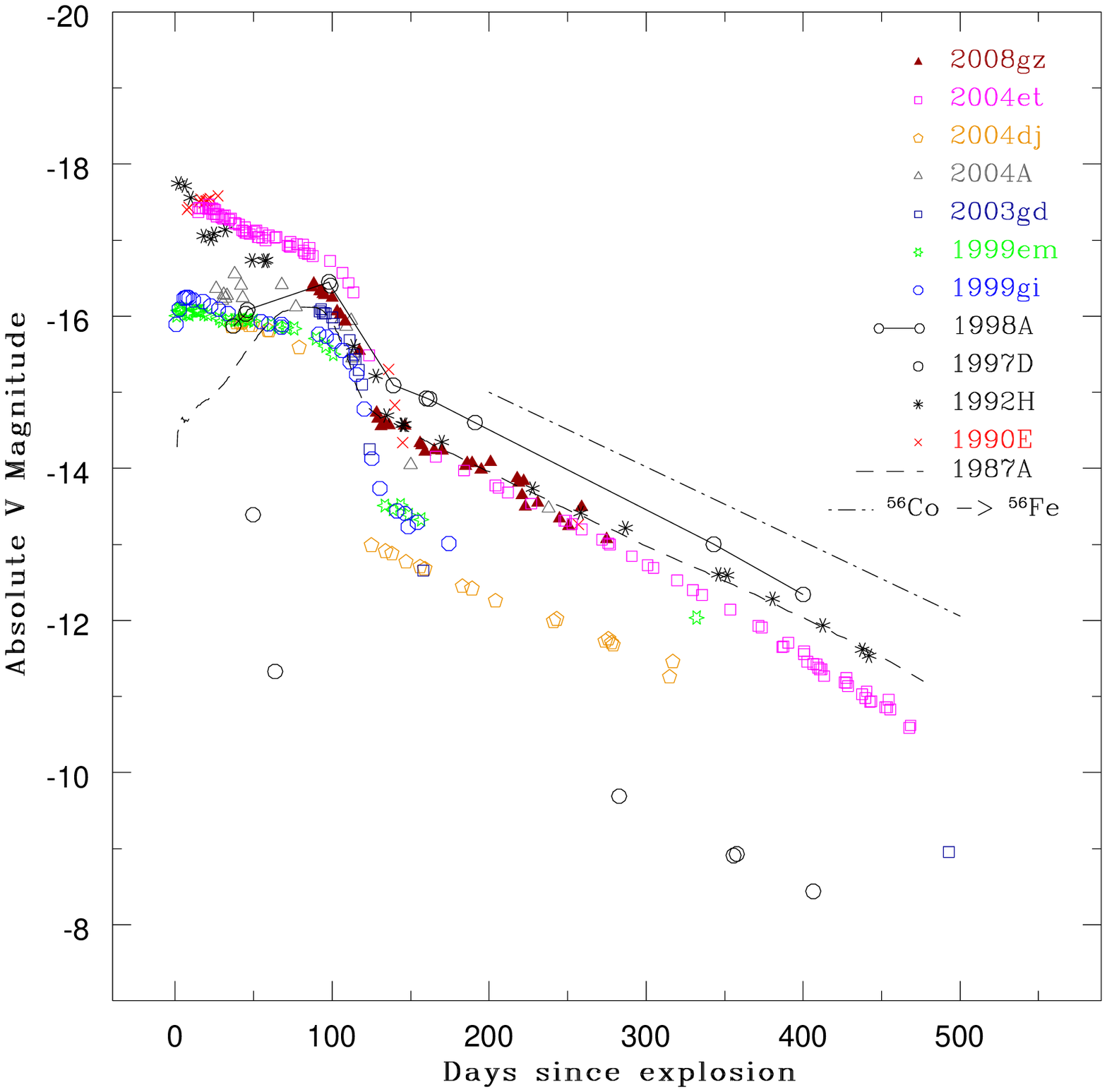}%
\caption{Comparison of absolute V-band light curve of SN 2008gz with other type
  IIP SNe, like SN 2004et, SN 2004dj, SN 2004A, SN 2003gd, SN 1999em, SN 1999gi, 
  SN 1998A, SN 1997D, SN 1992H, SN 1990E and SN 1987A. The magnitudes have been corrected
  for distance and readdening.}
\label{fig:abscur}
\end{figure}
 A detailed investigation of \sn\, indicates that it is a normal type IIP event showing 
 photometric and spectroscopic evolutions similar to archetypal SNe 2004et and 1999em. 
 \sn\, occurred in a highly inclined ($\Theta_{\rm inc} = 56.2 \degr$) host galaxy,
 within deprojected galactocentric radius of $0.27r_{\rm 25}$ (Table~\ref{tab:propgal}) 
 implying solar metallicity region similar to SNe 1999em and 2005cs, though its explosion
 and other properties were found to be similar to SN 2004et, which occurred in the outskirt of
 its host galaxy. Our narrow-band \ha\, photometry indicates that \sn\ was associated with 
 a star forming low luminosity \Hii\, region of the galaxy. Thus, the metallicty
 appear to have little effect on the explosion properties of core-collapse SNe.  

 In Fig.~\ref{fig:abscur} we show absolute V-band light curve of 
 well studied core-collapse events collected from literature \citep[see][for references]{misra07}.  
 The plateau luminosity ($M_V \sim -16.6$)\footnote{We estimate the value of mid plateau $M_V$ as 
 $-16.6\pm0.2$ mag, by considering an average decline rate of 0.006 mag d$^{-1}$ during the 
 plateau phase of SNe 1999em and 2004et, we found that the mid-plateau magnitude of \sn\, 
 was $\sim$ 0.2 mag brighter than that determined at +87d (\S\ref{sec:nick}). We assume $A_V$ of 0.214 
 mag.} of \sn\, is at the similar level as of peculiar SNe 1987A, 1998A and it lies in between
 to the brighter end of IIP SNe 1990E, 1992H, 2004et ($\sim -17$ mag) and the fainter end SN 1997D
 ($\sim -15$ mag). This shows that SN 2008gz is a normal event, both in energetics and nickel production.
 These values
 of plateau luminosity for IIP SNe are lower than $-17.6\pm0.6$ mag which was predicted using
 theoretical models calculated by \citet{hoflich01} for IIP SNe based on wide range of parameters 
 (explosion energy, metallicity, mass loss of progenitor). 

\begin{figure}
\centering
\includegraphics[scale = 0.42]{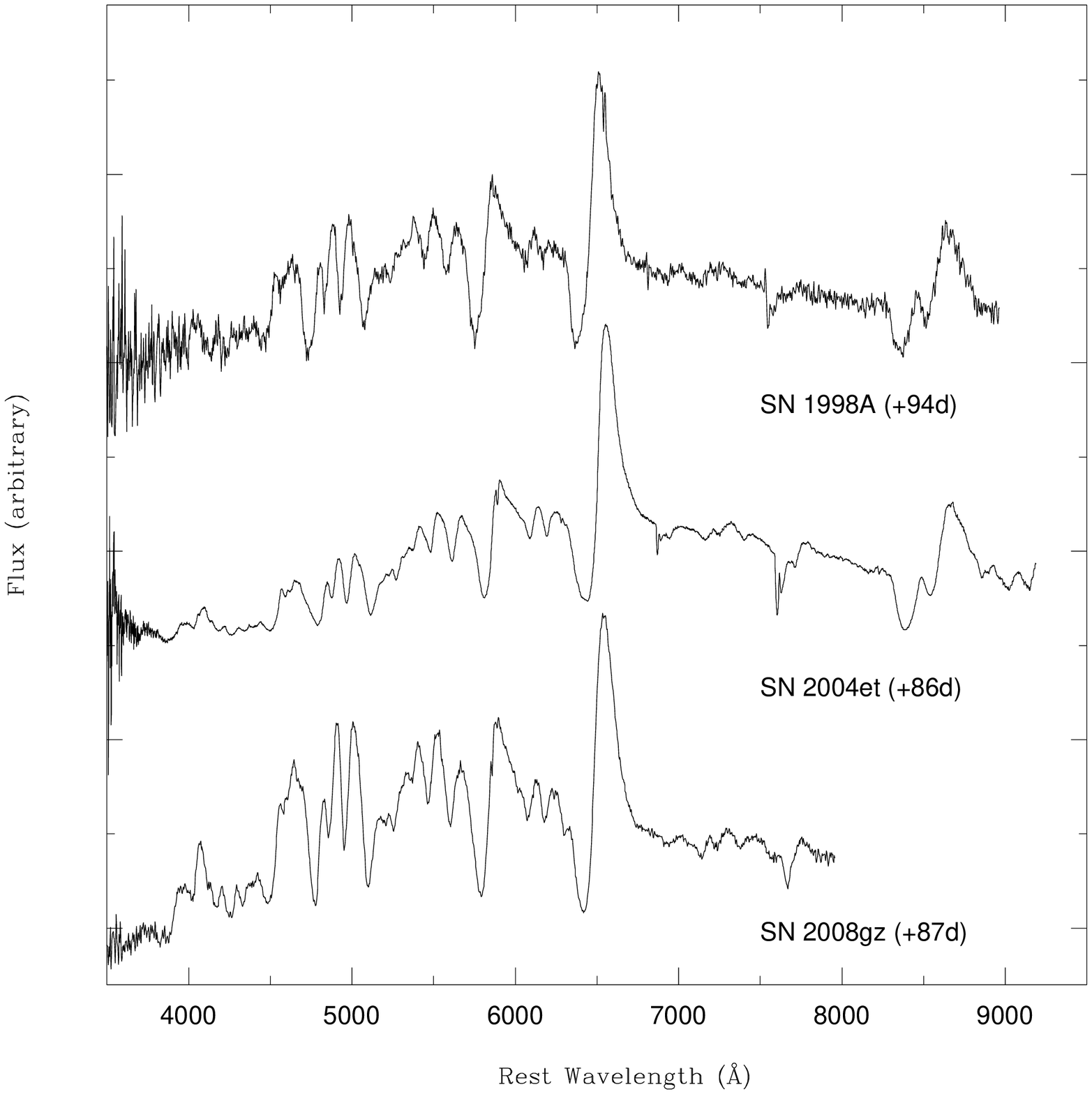}
\caption{Comparison of spectral features of SN 2008gz along with SN 2004et and SN 1998A}
\label{fig:speccomp}
\end{figure}

 \sn\, showed rarely observed 1.5 mag drop at $V$ from plateau to nebular phase
 and it had tail luminosity comparable (or higher) to SNe 2004et
 resulting in the synthesised \nickel\, mass in 
 range $0.05-0.1$\msun.  Large tail luminosity of SN 1998A indicates that the thermalization 
 process for 1998A was more efficient than \sn. Colour evolution of SN 2008gz has a similar trend 
 like normal IIP and peculiar SNe 1987A and 1998A. Expansion velocity of the \sn\, ejecta was 
 comparable to SN 2004et or higher, implying explosion energy of $\sim 2\times10^{51}$ erg. 
 Our calculation for progenitor mass (\S\ref{sec:proj}) of \sn\, favours the mass range 
 of $13-18$ M$_\odot$ (\ie, SNe 1999em, 1999gi, 2004dj, 2004et) than to the lower mass range
 of $8-18$ M$_\odot$ (\ie, SNe 1997D, 2004A, 2005cs). This progenitor mass grouping is also 
 favoured on the basis of radio luminosity  \citep{chevalier06}. 
 As an alternative scenario, \sn\ can be characterized as a peculiar event. Lack of data
 in first 1-2 months after the explosion and spectral similarity with SN 1998A
 \citep{benetti08} also indicates toward the possibility that \sn\ is a peculiar type II event.
 Figure~\ref{fig:speccomp}
 shows a comparison of SN 2008gz spectrum with SN 2004et and SN 1998A at comparable epoch. There
 are many similarities between spectral features of \sn\ and SN 1998A. On the other hand,
 low rates of peculiar type II SN, similarity in late light curve of \sn\ with type IIP events and
 similarity of some particular spectral features, like high velocity \ha\ line (\S\ref{sec:lines})
 of this event with that of SNe 2004et and 1999em also indicates that \sn\ would be a normal type IIP event.  


\section{summary} \label{sec:summary}
 We present $BVRI$ photometric and low resolution spectroscopic observations of a 
 supernova event SN 2008gz which occurred in an spiral arm and within half-light 
 radius of a nearby ($\sim$25 Mpc) galaxy NGC 3672. As the event was buried the
 galaxy light, we used template subtract technique to estimate the apparent
 magnitude of the event. We monitored the event for a period of $\sim$ 200 days. 
 We summarize our results as follows. 

 \begin{enumerate}

 \item 
  Photometric and spectral nature of the event is found similar to normal 
  type IIP SNe 2004et and 1999em. The event was discovered by about 82 days after 
  the burst and it has a plateau phase lasted $115\pm5$ days. We monitored the SN
  evolution from +87d to +275d. 

  \item 
  We estimate photospheric and H-envelop velocity by using both direct measurements of
  the absorption minimna of \Hi\, and \Feii\, lines and SYNOW modelling of the
  spectra. Here both values agree well within uncertainties. 
  We estimate the photospheric velocity of $\sim$4000 \kms at +87d, which is higher
  than that observed for well studied SN 2004et at similar epochs indicating
  explosion energy comparable to or higher than 2004et. Similar trend was
  also seen for the expansion velocity of H-envelopes. 
  
  \item
  Using pre-SN models of
  \cite{dessart10} and also by comparing explosion energies (derived by using
  hydrodynamical models) of well studied IIP SNe, we find that the \sn\, had
  explosion energy of $2-3\times10^{51}$ erg.s$^{-1}$. This estimation, coupled
  with the observed width of the forbidden \Oia\, line gives an upper limit
  for the main-sequence progenitor mass of 17\msun.

 \item 
  SN2008gz exhibits rarely observed drop of 1.5 mag within 30 days in V-band from plateau to 
  nebular phase, this is higher than the typically observed fall of 2-3 mag in 
  normal IIP SNe. Adopting A$_V$ = 0.21 mag,
  we could estimate mass of \nickel\, synthesized during the explosion as $0.05\pm0.01$\msun.

 \item
  Our \ha\, observation taken about 560 days after the explosion indicates 
  that the event took place in a low luminosity star forming arms, very close
  to a \Hii\, region. The emission kink of this \Hii\, region is also seen 
  in \ha\, line near zero velocity of the Doppler corrected spectra of SN. 

\end{enumerate}

\section*{Acknowledgments}
 We are thankful to the reviewer Stefano Valenti for his valuable comments,
 which have enriched the manuscript. 
 We thank all the observers at Aryabhatta Research Institute of Observational
 Sciences (ARIES) who provided their valuable time and support for the
 observations of this event. We are thankful to the observing
 staffs of 2-m IGO, 3.6-m TNG, 3.6-m NTT and 6-m BTA for their kind cooperation 
 in observation of SN 2008GZ. We also express our thanks to the observing staffs
 of Perth observatory for their kind support for this research work.
 This work was supported by the grant RNP 2.1.1.3483 of the Federal Agency of 
 Education of Russia. Timur A. Fatkhullin and Alexander S. Moskvitin were supported
 by the grant of the President of the Russian Federation (MK-405.2010.2).
 This work is partially based on observations made with the Italian Telescopio
 Nazionale Galileo (TNG) operated on the island of La Palma by the Fundación Galileo
 Galilei of the INAF (Istituto Nazionale di Astrofisica) at the Spanish Observatorio
 del Roque de los Muchachos of the Instituto de Astrofisica de Canarias. It is also
 partially based on observations collected at the European Southern Observatory,
 Chile under the program 083.D-0970(A). Stefano Benetti and Milena F Bufano are
 partially supported by the
 PRIN-INAF 2009 with the project ``Supernovae Variety and Nucleosynthesis Yields''
 This research has made use of data obtained through the High Energy Astrophysics
 Science Archive Research Center Online Service, provided by the NASA/Goddard
 Space Flight Center. We are indebted to the Indo-Russian (DST-RFBR) project No.
 RUSP-836 (RFBR-08-02:91314) for the completion of this research work. 


\clearpage
\setcounter{table}{1}
\begin{table}
  \caption{Journal of photometric observation of \sn.} 
  \label{tab:photlog}

  \begin{tabular}{c c c c c c c c cc} \hline
     UT Date&JD&Phase$^{a}$&$B$&$V$&$R$&$I$&Telescope$^{b}$& Seeing$^{c}$& Ellipticity$^{d}$ \\
     (yy/mm/dd)&2454000+&(day)&(s)&(s)&(s)&(s)& & (\arcsec)& \\ \hline
     2007/07/09 & 290.94 &$-$403& --           & 240          & 180          & 180          &LT & 2.5& 0.05 \\
     2008/11/10 & 781.48  &$+$87& 2$\times$300 & 2$\times$300 & 2$\times$250 & 2$\times$250 &ST & 2.4& 0.18 \\
     2008/11/11 & 782.50  &$+$88& 3$\times$300 & 300          & 2$\times$250 & 2$\times$250 &ST & 2.6& 0.33 \\
     2008/11/15 & 786.47  &$+$92& 250, 300     & 250, 300     & 250, 300     & 250, 300     &ST & 2.7& 0.11 \\
     2008/11/16 & 787.47  &$+$93& 2$\times$300 & 2$\times$300 & 250, 300     & 250, 300     &ST & 2.4& 0.13 \\
     2008/11/17 & 788.48  &$+$94& 2$\times$300 & 2$\times$300 & 2$\times$300 & 2$\times$300 &ST & 2.4& 0.10 \\
     2008/11/18 & 789.46  &$+$95& 2$\times$300 & 2$\times$300 & 2$\times$300 & 2$\times$250 &ST & 3.3& 0.11 \\
     2008/11/23 & 794.47 &$+$100& 2$\times$300 & 2$\times$300 & 2$\times$300 & 2$\times$300 &ST & 3.5& 0.05 \\
     2008/11/26 & 797.47 &$+$103& 240          & 150          & 60           & 60           &ST & 2.7& 0.05 \\
     2008/11/28 & 799.45 &$+$105& 2$\times$300 & 2$\times$300 & 2$\times$300 & 2$\times$300 &ST & 3.0& 0.13 \\
     2008/12/01 & 802.49 &$+$108& 2$\times$300 & 2$\times$300 & 2$\times$300 & 2$\times$300 &ST & 2.5& 0.18 \\
     2008/12/10 & 811.49 &$+$117& 2$\times$300 & 2$\times$300 & 2$\times$300 & 2$\times$300 &ST & 2.9& 0.21 \\
     2008/12/21 & 822.49 &$+$128& 2$\times$300 & 2$\times$300 & 2$\times$300 & 2$\times$300 &ST & 2.6& 0.14 \\
     2008/12/22 & 823.46 &$+$129& 2$\times$300 & 2$\times$300 & 2$\times$300 & 2$\times$300 &ST & 2.3& 0.20 \\
     2008/12/24 & 825.51 &$+$131& 300          & 300          & 300          & 250          &ST & 3.6& 0.19 \\
     2008/12/26 & 827.52 &$+$133& 2$\times$300 & 300          & 300          & 300          &ST & 3.1& 0.24 \\
     2008/12/28 & 829.45 &$+$135& --           & 300          & 300          & 300          &ST & 2.7& 0.09 \\
     2008/12/29 & 830.49 &$+$136& 2$\times$300 & 300          & 300          & 2$\times$300 &ST & 3.2& 0.05 \\
     2009/01/01 & 833.51 &$+$139& --           & 300          & 300          & 300          &ST & 2.7& 0.12 \\
     2009/01/08 & 840.48 &$+$146& 2$\times$300 & 300          & 300          & 2$\times$300 &ST & 3.1& 0.25 \\
     2009/01/18 & 850.36 &$+$156& 2$\times$300 & 2$\times$300 & 2$\times$300 & 2$\times$300 &ST & 3.1& 0.16 \\
     2009/01/19 & 851.32 &$+$157& 3$\times$300 & 2$\times$300 & 2$\times$300 & 2$\times$300 &ST & 3.1& 0.11 \\
     2009/01/21 & 853.32 &$+$159& 2$\times$300 & 2$\times$300 & 2$\times$300 & 2$\times$300 &ST & 3.2& 0.08 \\
     2009/01/23 & 855.51 &$+$161& 2$\times$300 & 2$\times$300 & 2$\times$300 &              &ST & 3.1& 0.36 \\
     2009/01/27 & 859.36 &$+$165& 2$\times$300 & 2$\times$300 & 2$\times$300 & 2$\times$300 &ST & 2.2& 0.18 \\
     2009/02/01 & 864.35 &$+$170& 2$\times$300 & 2$\times$300 & 2$\times$300 & 2$\times$300 &ST & 2.5& 0.11 \\
     2009/02/16 & 879.42 &$+$185& 2$\times$200 & 2$\times$200 & 2$\times$250 & 2$\times$250 &ST & 2.4& 0.42 \\
     2009/02/17 & 880.27 &$+$186& 2$\times$250 & 2$\times$250 & 2$\times$200 & 2$\times$200 &ST & 2.9& 0.17 \\
     2009/02/20 & 883.34 &$+$189& 2$\times$300 & 2$\times$300 & 2$\times$300 & 2$\times$300 &ST & 3.2& 0.24 \\
     2009/02/26 & 889.27 &$+$195& 2$\times$300 & 3$\times$300 & 2$\times$300 & 2$\times$300 &ST & 3.6& 0.14 \\
     2009/03/04 & 895.23 &$+$201& 2$\times$300 & 2$\times$300 & 2$\times$300 & 2$\times$300 &ST & 3.0& 0.18 \\
     2009/03/21 & 912.33 &$+$218& 1200         & 600          & 300          & 300          &IGO& 1.6& 0.10 \\
     2009/03/22 & 913.24 &$+$219& 1200         & 600          & 300          & 300          &IGO& 1.6& 0.13 \\
     2009/03/24 & 915.31 &$+$221& 1200         & 600          & 300          & 300          &IGO& 1.5& 0.03 \\
     2009/03/25 & 916.30 &$+$222& 1200         & 600          & 300          & 300          &IGO& 1.4& 0.05 \\
     2009/03/26 & 917.32 &$+$223& 2$\times$300 & 2$\times$300 & 2$\times$300 & 2$\times$300 &ST & 3.5& 0.14 \\
     2009/04/03 & 925.20 &$+$231& 2$\times$300 & 2$\times$300 & 2$\times$300 & 3$\times$300 &ST & 2.8& 0.24 \\
     2009/04/12 & 934.28 &$+$240& --           & 2$\times$300 & 2$\times$300 & 2$\times$300 &ST & 2.7& 0.43 \\
     2009/04/17 & 939.23 &$+$245& 2$\times$300 & 2$\times$250 & 2$\times$300 & 2$\times$250 &ST & 2.8& 0.16 \\
     2009/04/23 & 945.16 &$+$251& 3$\times$300 & 2$\times$300 & 2$\times$300 & 2$\times$250 &ST & 3.1& 0.16 \\
     2009/05/01 & 953.15 &$+$259& 2$\times$300 & 2$\times$300 & 300          & 300          &ST & 2.5& 0.38 \\
     2009/05/17 & 969.11 &$+$275& 3$\times$120 & 3$\times$60  & 60           & 60           &NTT& 0.9& 0.08 \\
     2009/11/19 & 1155.94&$+$462& 2$\times$250 & --           & --           & --           &ST & -- & --   \\
     2010/02/13 & 1241.34&$+$547& --           & 3$\times$300 & 3$\times$300 & 3$\times$300 &ST & 2.5& 0.13 \\  
     2010/02/14 & 1242.35&$+$548& 4$\times$300 & 2$\times$300 & 2$\times$300 & 2$\times$300 &ST & 2.1& 0.12 \\  
     \hline               
  \end{tabular}
  \newline
  $^{a}$ With reference to the time of explosion JD 2454694.0.\\ 
  $^{b}$ LT : 0.6 m Lowell Telescope, Perth Observatory, UK; 
  ST : 1 m Sampurnanand Telescope, ARIES, India; IGO : 2 m IUCAA 
  Girawali Observatory, India; NTT : 3.6 m New Technology Telescope, ESO, Chile\\
  $^{c}$ FWHM of stellar PSF at $V$ band\\ 
  $^{d}$ Ellipticity is estimated using {\sc imexamine} task of IRAF\\ 
 
\end{table}


\setcounter{figure}{2}
\clearpage
\begin{figure*}
\centering
\includegraphics[width=17cm]{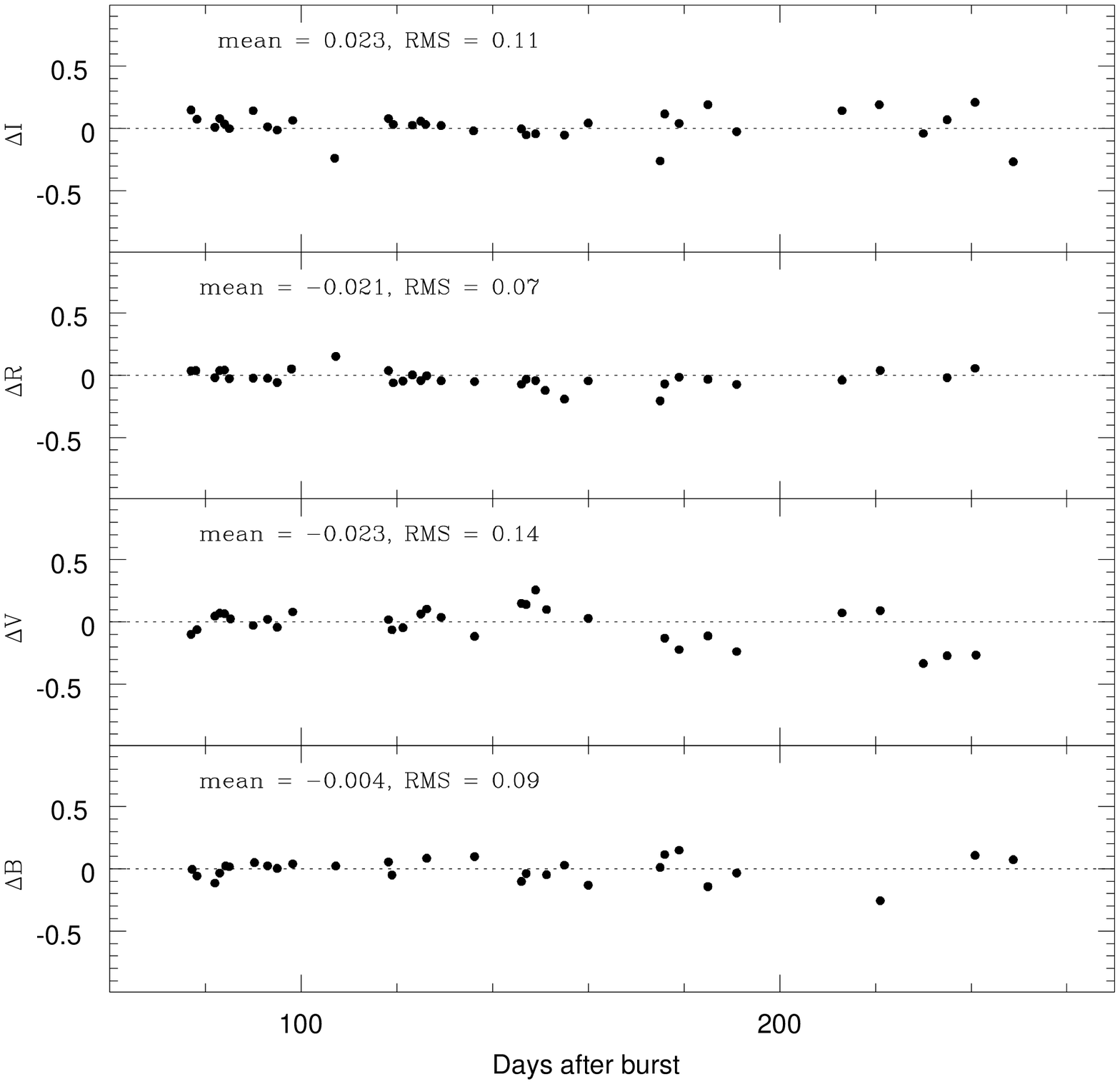}
\caption{Comparison of ISIS derived magnitudes and ours for $BVRI$.}
\label{fig:diffplot}
\end{figure*}

\setcounter{figure}{7}
\begin{figure*}
\centering
\includegraphics[scale=0.8]{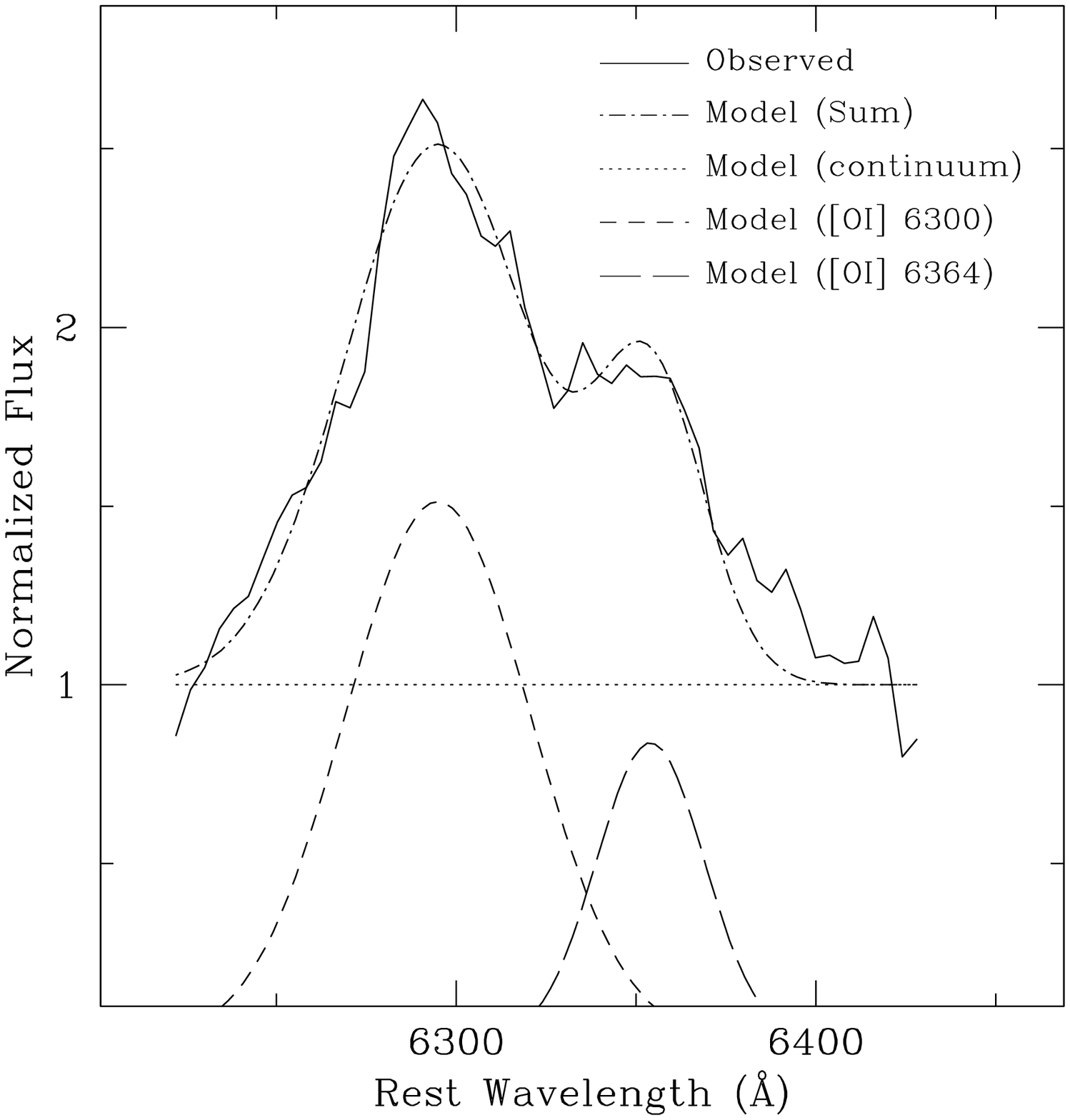}%
\caption{Two component Gaussian fit of \Oia\ 6300, 6364 \AA\ emission lines. 
  Measurements of their relative intensities quantifies, whether the corresponding
  line emitting region is optically thick or thin.}
\label{fig:specoii}
\end{figure*}

\setcounter{figure}{11}
\begin{figure*}
\centering
\includegraphics[width=17cm]{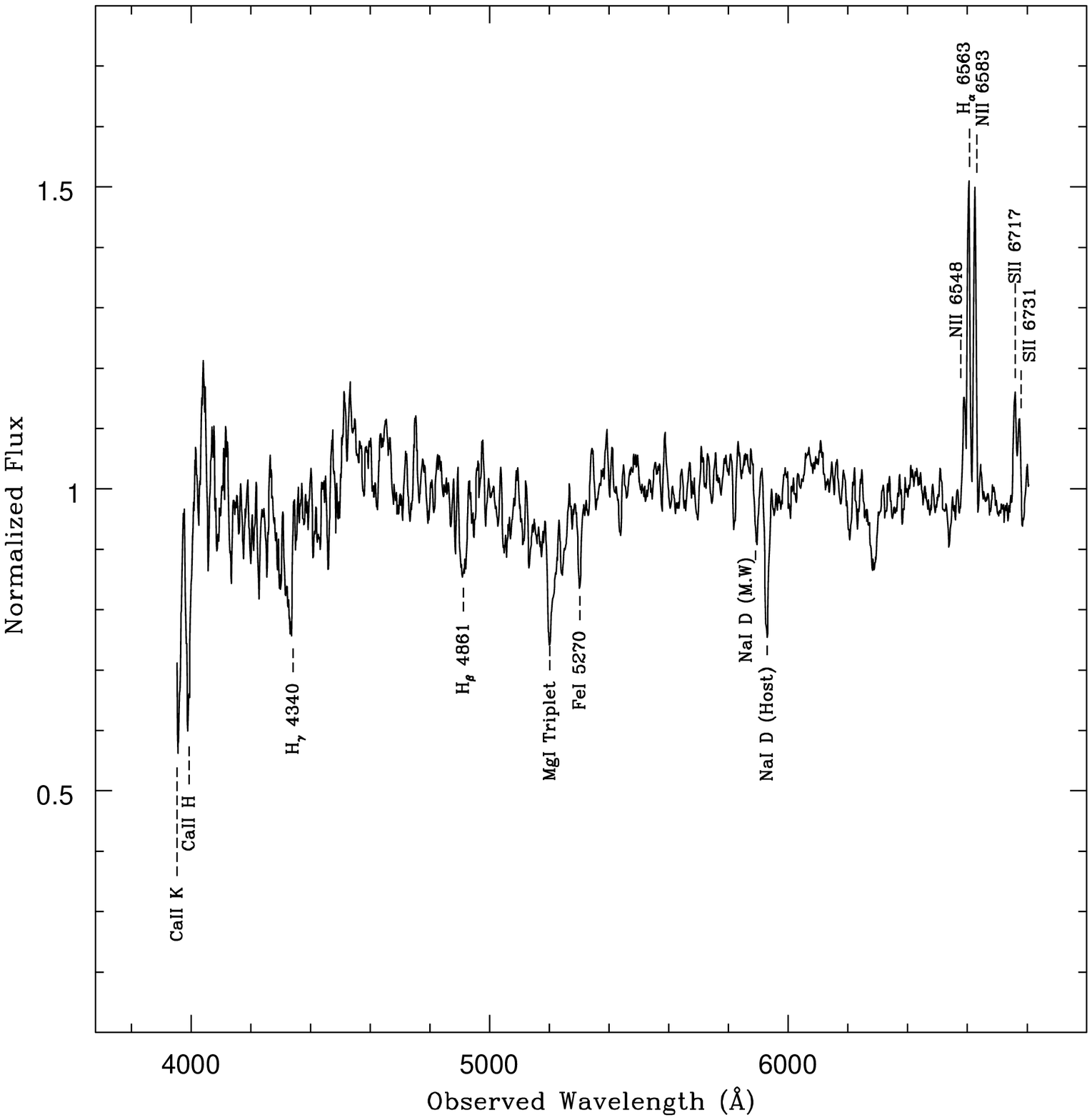}%
\caption{Spectrum of nucleus of the host galaxy NGC 3672 taken with 2-m IGO. The spectrum
        shows \ha, \Nii\,(6548 \AA\ and 6583\AA), and \Sii\,(6717\AA and 6730\AA) lines in emission
        (similar to Sc spiral galaxies), which can arise from a gaseous component heated
        by AGN, post-AGB stars, shocks or cooling flows. The \Caii\,(K and H), \hg, \hb, \Mgi\,T,
        \Fei\,(5270\AA), \Nai\,D due to Milky Way and the host are seen in absorption.}
\label{fig:specgal}
\end{figure*}

\setcounter{figure}{16}
\begin{figure*}
\centering
\includegraphics[scale=0.7]{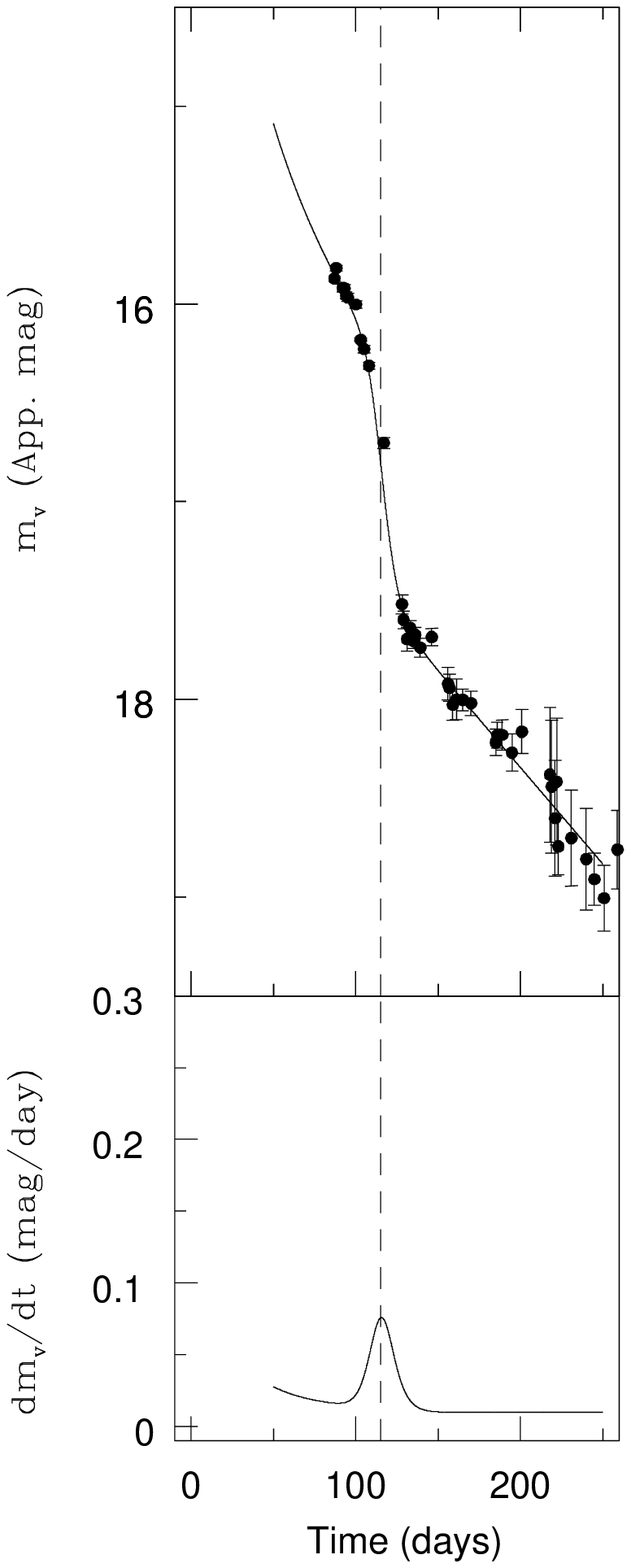}
\caption{Steepness parameter estimation for \sn.}
\label{fig:steep}
\end{figure*}

%

\end{document}